\RequirePackage{fix-cm}
\documentclass[smallextended]{svjour3}       
\smartqed  

\usepackage[ruled]{algorithm2e}

\usepackage{commath}

\usepackage{booktabs}
\usepackage{multirow}
\usepackage[flushleft]{threeparttable}

\usepackage{float}

\usepackage{empheq}
\newlength\fsep
\newsavebox\widebox
\newenvironment{mathbox}
{\par\vskip\fsep\noindent%
	\begin{lrbox}{\widebox}%
		\begin{minipage}{\textwidth-\fsep}%
		}{\vskip\fsep\end{minipage}\end{lrbox}
	\framebox{\usebox\widebox}%
}

\usepackage{graphicx}
\usepackage{wrapfig}

\usepackage{caption}
\usepackage{subcaption}
\usepackage{listings}
\usepackage{textcomp}

\SetAlFnt{\small}
\SetAlCapFnt{\small}
\SetAlCapNameFnt{\small}
\SetAlCapHSkip{0pt}
\IncMargin{-\parindent}

\begin{document}

\title{CHAOS: A Parallelization Scheme for Training Convolutional Neural Networks on Intel Xeon Phi 
}

\titlerunning{CHAOS: A Parallelization Scheme for Training Convolutional Neural Networks}        

\author{Andr\'{e} Viebke         \and
        Suejb Memeti \and
        Sabri Pllana \and
        Ajith Abraham
}

\authorrunning{A. Viebke, S. Memeti, S. Pllana, and A. Abraham} 

\institute{ A. Viebke \and S. Memeti \and S. Pllana \at
              Linnaeus University \\
              Address: Department of Computer Science, 351 95 V\"{a}xj\"{o}, Sweden\\
              A. Viebke: \email{av22cj@student.lnu.se}\\
              S. Memeti: \email{suejb.memeti@lnu.se}\\
              S. Pllana: \email{sabri.pllana@lnu.se}
            \and
            A. Abraham \at
              Machine Intelligence Research Labs (MIR Labs) \\
              Address:  1, 3rd Street NW, P.O. Box 2259 Auburn, Washington 98071, USA \\
              \email{ajith.abraham@ieee.org}
}

\date{Received: date / Accepted: date}

\maketitle

\begin{abstract}
Deep learning is an important component of big-data analytic tools and intelligent applications, such as, self-driving cars, computer vision, speech recognition, or precision medicine. However, the training process is computationally intensive, and often requires a large amount of time if performed sequentially. Modern parallel computing systems provide the capability to reduce the required training time of deep neural networks.
In this paper, we present our parallelization scheme for training convolutional neural networks (CNN) named Controlled Hogwild with Arbitrary Order of Synchronization (CHAOS). Major features of CHAOS include the support for thread and vector parallelism, non-instant updates of weight parameters during back-propagation without a significant delay, and implicit synchronization in arbitrary order. CHAOS is tailored for parallel computing systems that are accelerated with the Intel Xeon Phi. We evaluate our parallelization approach empirically using measurement techniques and performance modeling for various numbers of threads and CNN architectures. Experimental results for the MNIST dataset of handwritten digits using the total number of threads on the Xeon Phi show speedups of up to $103\times$ compared to the execution on one thread of the Xeon Phi, $14\times$ compared to the sequential execution on Intel Xeon E5, and $58\times$ compared to the sequential execution on Intel Core i5. 


\end{abstract}

\section{Introduction}

Traditionally engineers developed applications by specifying computer instructions that determined the application behavior. Nowadays engineers focus on developing and implementing sophisticated deep learning models that can learn to solve complex problems. Moreover, deep learning algorithms \cite{lecun2015nature} can learn from their own experience rather than that of the engineer. 

Many private and public organizations are collecting huge amounts of data that may contain useful information from which valuable knowledge may be derived. With the pervasiveness of the Internet of Things the amount of available data is getting much larger \cite{Hsu2014}. Deep learning is a useful tool for analyzing and learning from massive amounts of data (also known as Big Data) that may be unlabeled and unstructured \cite{strawn2016,sharma2014,najafabadi2015}. Deep learning algorithms can be found in many modern applications \cite{siri,taigman2014deepface,hadsell2009learning,wu2014deep,szegedy2015going,fox2014,jiang2016,zhang2012}, such as, voice recognition, face recognition, autonomous cars, classification of liver diseases and breast cancer, computer vision, or social media. 

A Convolutional Neural Network (CNN) is a variant of a Deep Neural Network (DNN) \cite{110_deeplearning_lenet}. Inspired by the visual cortex of animals, CNNs are applied to state-of-the-art applications, including computer vision and speech recognition \cite{deng2014deep}. However, supervised training of CNNs is computationally demanding and time consuming, and in many cases, several weeks are required to complete a training session. Often applications are tested with different parameters, and each test requires a full session of training. 

Multi-core processors~\cite{Williams:2009} and in particular many-core~\cite{benkner11} processing architectures, such as the NVIDIA Graphical Processing Unit (GPU) \cite{nvidia_gpu} or the Intel Xeon Phi \cite{chrysos2012intel} co-processor, provide processing capabilities that may be used to significantly speed-up the training of CNNs. While existing research \cite{cirecsan2011high,vrtanoski2012pattern,szegedy2015going,yadan2013multi,sainath2015deep} has addressed extensively the training of CNNs using GPUs, so far not much attention is given to the Intel Xeon Phi co-processor. Beside the performance capabilities, the Xeon Phi deserves our attention because of programmability \cite{pllana09} and portability \cite{KesslerDTNRDBTP12}. 

In this paper, we present our parallelization scheme for training convolutional neural networks, named Controlled Hogwild with Arbitrary Order of Synchronization (CHAOS). CHAOS is tailored for the Intel Xeon Phi co-processor and exploits both the thread- and SIMD-level parallelism. The thread-level parallelism is used to distribute the work across the available threads, whereas SIMD parallelism is used to compute the partial derivatives and weight gradients in convolutional layer. Empirical evaluation of CHAOS is performed on an Intel Xeon Phi 7120 co-processor. For experimentation, we use various number of threads, different CNNs architectures, and the MNIST dataset of handwritten digits \cite{lecun2010mnist}. Experimental evaluation results show that using the total number of available threads on the Intel Xeon Phi we can achieve speedups of up to $103\times$ compared to the execution on one thread of the Xeon Phi, $14\times$ compared to the sequential execution on Intel Xeon E5, and $58\times$ compared to the sequential execution on Intel Core i5. The error rates of the parallel execution are comparable to the sequential one. Furthermore, we use performance prediction to study the performance behavior of our parallel solution for training CNNs for numbers of cores that go beyond the generation of the Intel Xeon Phi that was used in this paper. The main contributions of this paper include:

\begin{itemize}
	\item design and implementation of CHAOS parallelization scheme for training CNNs on the Intel Xeon Phi,
	\item performance modeling of our parallel solution for training CNNs on the Intel Xeon Phi,
	\item measurement-based empirical evaluation of CHAOS parallelization scheme,
	\item model-based performance evaluation for future architectures of the Intel Xeon Phi. 
\end{itemize}

The rest of the paper is organized as follows. We discuss the related work in Section \ref{sec:related-work}. Section \ref{sec:background} provides background information on CNNs and the Intel Xeon Phi many-core architecture. Section \ref{sec:chaos} discusses the design and implementation aspects of our parallelization scheme. The experimental evaluation of our approach is presented in Section \ref{sec:evaluation}. We summarize the paper in Section~\ref{sec:conclusion}.

\section{Related Work}
\label{sec:related-work}

In comparison to related work that target GPUs, the work related to machine learning for Intel Xeon Phi is sparse. In this section, we describe machine learning approaches that target the Intel Xeon Phi co-processor, and thereafter we discuss CNN solutions for GPUs and contrast them to our CHAOS implementation. 

\subsection{Machine Learning targeting Intel Xeon Phi}
\label{sec:rw-xphi}

In this section, we discuss existing work for Support Vector Machines (SVMs), Restricted Boltzmann Machines (RBMs), sparse auto encoders and the Brain-State-in-a-Box (BSB) model.

You et al. \cite{you2014mic} present a library for parallel Support Vector Machines, MIC-SVM, which facilitates the use of SVMs on many- and multi-core architectures including Intel Xeon Phi. Experiments performed on several known datasets showed up to 84x speed up on the Intel Xeon Phi compared to the sequential execution of LIBSVM \cite{chang2011libsvm}. In comparison to their work, we target deep learning.

Jin et al. \cite{JinWGYH14} perform the training of sparse auto encoders and restricted Boltzmann machines on the Intel Xeon Phi 5110p. The authors reported a speed up factor of $7 - 10\times$ times compared to the Xeon E5620 CPU and more than $300\times$ times compared to the un-optimized version executed on one thread on the co-processor. Their work targets unsupervised deep learning of restricted Boltzmann machines and sparse auto encoders, whereas we target supervised deep learning of CNNs.

The performance gain on Intel Xeon Phi 7110p for a model called Brain-State-in-a-Box (BSB) used for text recognition is studied by Ahmed et al. in \cite{ahmed2014accelerating}. The authors report about two-fold speedup for the co-processor compared to a CPU with 16 cores when parallelizing the algorithm. While both approaches target Intel Xeon Phi, our work addresses training of CNNs on the MNIST dataset.

\subsection{Related Work Targeting CNNs}
\label{sec:rw-cnn}

In this section, we will discuss CNNs solutions for GPUs in the context of computer vision (image classification). Work related to MNIST \cite{lecun2010mnist} dataset is of most interest, also NORB \cite{lecun2004learning} and CIFAR 10 \cite{krizhevsky2009learning} is considered. Additionally, work done in speech recognition and document processing is briefly addressed. We conclude this section by contrasting the presented related work with our CHAOS parallelization scheme. 

Work presented by Cire{\c{s}}an et al. \cite{cirecsan2011high} target a CNN implementation raising the bars for the CIFAR10 (19.51\% error rate), NORB (2.53\% error rate) and MNIST (0.35\% error rate) datasets. The training was performed on GPUs (Nvidia GTX 480 and GTX 580) where the authors managed to decrease the training time severely - up to $60\times$ compared to sequential execution on a CPU - and decrease the error rates to an, at the time, state-of-the-art accuracy level.

Later, Cire{\c{s}an et al. \cite{cirecsan2012multi} presented their multi-column deep neural network for classification of traffic sings. The results show that the model performed almost human-like (humans\textquotesingle error rate about 0.20\%) on the MNIST dataset, achieving a best error rate of 0.23\%. The authors trained the network on a GPU.
	
Vrtanoski et al. \cite{vrtanoski2012pattern} use OpenCL for parallelization of the back-propagation algorithm for pattern recognition. They showed a significant cost reduction, a maximum speedup of $25.8\times$ was achieved on an ATI 5870 GPU compared to a Xeon W3530 CPU when training the model on the MNIST dataset.
	
The ImageNet challenge aims to evaluate algorithms for large-scale object detection and image classification based on the ImageNet dataset. Krizhevsky et al. \cite{krizhevsky2012imagenet} joined the challenge and reduced the error rate of the test set to 15.3\% from the second best 26.2\% using a CNN with 5 convolutional layers. For the experiments, two GPUs (Nvidia GTX 580) were used only communicating in certain layers. The training lasted for 5 to 6 days.
	
In a later challenge, ILSVRC 2014, a team from Google entered the competition with GoogleNet, a 22-layer deep CNN and won the classification challenge with a 6.67\% error rate. The training was carried out on CPUs. The authors state that the network could be trained on GPUs within a week, illuminating the limited amount of memory to be one of the major concerns \cite{szegedy2015going}.
	
Yadan et al. \cite{yadan2013multi} used multiple GPUs to train CNNs on the ImageNet dataset using both data- and model-parallelism, i.e. either the input space is divided into mini-batches where each GPU train its own batch (data parallelism) or the GPUs train one sample together (model parallelism). There is no direct comparison with the training time on CPU, however, using 4 GPUs (Nvidia Titan) and model- and data-parallelism, the network was trained for 4.8 days.
	
Song et al. \cite{song2014deep} constructed a CNN to recognize face expressions and developed a smart-phone app in which the user can capture a picture and send it to a server hosting the network. The network, predicts a face expression and sends the result back to the user. With the help of GPUs (Nvidia Titan), the network was trained in a couple of hours on the ImageNet dataset.
	
Scherer et al. \cite{scherer2010accelerating} accelerated the large-scale neural networks with parallel GPUs. Experiments with the NORB dataset on an Nvidia GTX 285 GPU showed a maximal speedup of $115\times$ compared to a CPU implementation (Core i7 940). After training the network for 360 epochs, an error rate of 8.6\% was achieved.
	
Cire{\c{s}}an et al. \cite{cirecsan2011committee} combined multiple CNNs to classify German traffic signs and achieved a 99.15\% recognition rate (0.85 \% error rate). The training was performed using an Intel Core i7 and 4 GPUs (2 x GTX 480 and 2 x GTX 580).
	
More recently Abadi et al. \cite{tensorflow2015whitepaper} presented TensorFlow, a system for expressing and executing machine learning algorithms including training deep neural network models.
	
Researchers have also found CNNs successful for speech tasks. Large vocabulary continuous speech recognition deals with translation of continuous speech for languages with large vocabularies. Sainath et al. \cite{sainath2015deep} investigated the advantages of CNNs performing speech recognition tasks and compared the results with previous DNN approaches. Results indicated on a 12-14\% relative improvement of word error rates compared to a DNN trained on GPUs.
	
Chellapilla et al. \cite{chellapilla2006high} investigated GPUs (Nvidia Geforce 7800 Ultra) for document processing on the MNIST dataset and achieved a $4.11\times$ speed up compared to the sequential execution a Intel Pentium 4 CPU running at 2.5 GHz clock frequency.

In contrast to CHAOS, these studies target training of CNNs using GPUs, whereas our approach addresses training of CNNs on the MNIST dataset using the Intel Xeon Phi co-processor. While there are several review papers (such as, \cite{bahrampour2015comparative,shi2016benchmarking,tabik2017snapshot}) and on-line articles (such as, \cite{murphy2016review}) that compare existing frameworks for parallelization of training CNN architectures, we focus on detailed analysis of our proposed parallelization approach using measurement techniques and performance modeling. We compare the performance improvement achieved with CHAOS parallelization scheme to the sequential version executed on Intel Xeon Phi, Intel Xeon E5 and Intel Core i5 processor.

\section{Background}
\label{sec:background}

In this section, we first provide some background information related to the neural networks focusing on convolutional neural networks, and thereafter we provide some information about the architecture of the Intel Xeon Phi.

\subsection{Neural Networks}
\label{sec:nn}

A Convolutional Neural Network is a variant of a Deep Neural Network, which introduces two additional layer types: \emph{convolutional layers} and \emph{pooling layers}. The mammal visual processing system is hierarchical (deep) in nature. Higher level features are abstractions of lower level ones. E.g. to understand speech, waveforms are translated through several layers until reaching a linguistic level. A similar analogy can be drawn for images, where edges and corners are lower level abstractions translated into more spatial patterns on higher levels. Moreover, it is also known that the animal cortex consists of both simple and complex cells firing on certain visual inputs in their receptive fields. Simple cells detect edge-like patterns whereas complex cells are locally invariant, spanning larger receptive fields. These are the very fundamental properties of the animal brain inspiring DNNs and CNNs. 

In this section, we first describe the DNNs and the Forward- and Back-propagation, thereafter we introduce the CNNs.

\subsubsection{Deep Neural Networks}
\label{sec:dnn}

The architecture of a DNN consists of multiple layers of neurons. Neurons are connected to each other through edges (weights). The network can simply be thought of as a weighted graph; a directed acyclic graph represents a feed-forward network. The depth and breadth of the network differs as may the layer types. Regardless of the depth, a network has at least one input and one output layer. A neuron has a set of incoming weights, which have corresponding outgoing edges attached to neurons in the previous layer. Also, a bias term is used at each layer as an intercept term. The goal of the learning process is to adjust the network weights and find a global minimum by reducing the overall error, i.e. the deviation between the predicted and the desired outcome of all the samples. The resulting weight parameters can thereafter be used to make predictions of unseen inputs \cite{ng2011ufldl}.

\subsubsection{Forward Propagation}
\label{sec:fw-propagation}

DNNs can make predictions by forward propagating an input through the network. Forward propagation proceeds by performing calculations at each layer until reaching the output layer, which contains a vector representing the prediction. For example, in image classification problems, the output layer contains the prediction score that indicates the likelihood that an image belongs to a category \cite{agibansky,ng2011ufldl}. 

The forward propagation starts from a given input layer, then at each layer the activation for a neuron is activated using the equation $y^l_i = \sigma(x^l_i) + I^l_i$ where $y^l_i$ is the output value of neuron $i$ at layer $l$, $x^l_i$ is the input value of the same neuron, and $\sigma$ (sigmoid) is the activation function. $I^l_i$ is used for the input layer when there is no previous layer. The goal of the activation function is to return a normalized value (\textit{sigmoid} return [0,1] and \textit{tanh} is used in cases where the desired return values are [-1,1]). 
The input $x^l_i$ can be calculated as $x^l_i = \sum_j(w^{l}_{ji}y^{l-1}_j)$ where $w^{l}_{ji}$ denotes the weight between neuron $i$ in the current layer $l$, and $j$ in the previous layer, and $y^{l-1}_j$ the output of the $j$th neuron at the previous layer. This process is repeated until reaching the output layer. At the output layer, it is common to apply a soft max function, or similar, to squash the output vector and hence derive the prediction. 

\subsubsection{Back-Propagation}
\label{sec:bw-propagation}

Back-propagation is the process of propagating errors, i.e. the loss calculated as the deviation between the predicted and the desired output, backward in the network, by adjusting the weights at each layer. The error and partial derivatives $\delta^l_i$ are calculated at the output layer based on the predicted values from forward propagation and the labeled value (the correct value). At each layer, the relative error of each neuron is calculated and the weight parameters are updated based on how much the neuron participated in the faulty prediction. The equation: $\dfrac{\delta E}{\delta y^l_i} = \sum{w^l_{ij} \dfrac{\delta E}{\delta x^{l+1}_j}}$ denotes that the partial derivative of neuron $i$ at the current layer $l$ is the sum of the derivatives of connected neurons at the next layer multiplied with the weights, assuming $w^l$ denotes the weights between the maps.
Additionally, a decay is commonly used to control the impact of the updates, which is omitted in the above calculations. More concretely, the algorithm can be thought of as updating the layer\textquotesingle s weights based on "how much it was responsible for the errors in the output" \cite{agibansky,ng2011ufldl}.

\subsubsection{Convolutional Neural Networks}
\label{sec:cnn}

A Convolutional Neural Network is a multi-layer model constructed to learn various levels of representations where higher level representations are described based on the lower level ones \cite{schmidhuber2015deep}. It is a variant of deep neural network that introduces two new layer types: \emph{convolutional} and \emph{pooling} layers.

The convolutional layer consists of several feature maps where neurons in each map connect to a grid of neurons in maps in the previous layer through overlapping kernels. The kernels are tiled to cover the whole input space. The approach is inspired by the receptive fields of the mammal visual cortex. All neurons of a map extract the same features from a map in the previous layer as they share the same set of weights.

Pooling layers intervene convolutional layers and have shown to lead to faster convergence. Each neuron in a pooling layer outputs the (maximum/average) value of a partition of neurons in the previous layer, and hence only activates if the underlying grid contains the sought feature. Besides from lowering the computational load, it also enables position invariance and down samples the input by a factor relative to the kernel size \cite{lecun1998gradient}.

Figure \ref{fig:LeNet} shows LeNet-5 that is an example of a Convolutional Neural Network. Each layer of convolution and pooling (that is a specific method of sub-sampling used in LeNet) comprise several feature maps. Neurons in the feature map cover different sub-fields of the neurons from the previous layer. All neurons in a map share the same weight parameters, therefore they extract the same features from different parts of the input from the previous layers. 

\begin{figure}[bt]
	\center
	\includegraphics[width=0.9\linewidth]{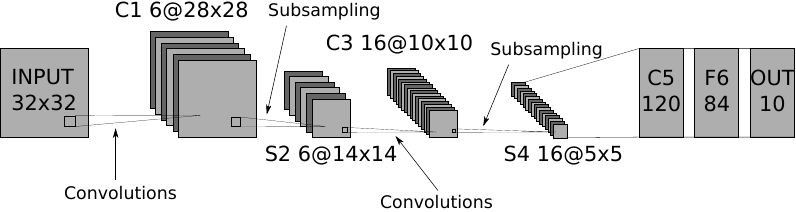}
	\caption{The LeNet-5 architecture.}
	\label{fig:LeNet}
\end{figure}

CNNs are commonly constructed similarly to the LeNet-5, beginning with an input layer, followed by several convolutional/pooling combinations, ending with a fully connected layer and an output layer \cite{lecun1998gradient}. Recent networks are much deeper and/or wider, for instance, the GoogleNet \cite{szegedy2015going} consists of 22 layers.

Various implementations target the Convolutional Neural Networks, such as: \emph{EbLearn} at New York University and \emph{Caffe} at Berkeley. As a basis for our work we selected a project developed by Cire{\c{s}}an \cite{dciresan}. This implementation targets the MNIST dataset of handwritten digits, and has the possibility to dynamically configure the definition of layers, the activation function and the connection types using a configuration file.

\subsection{Parallel Systems accelerated with Intel\textregistered Xeon Phi\texttrademark}
\label{sec:xphi}

\begin{figure}[bt]
	\begin{center}
		\includegraphics[width=0.5\textwidth]{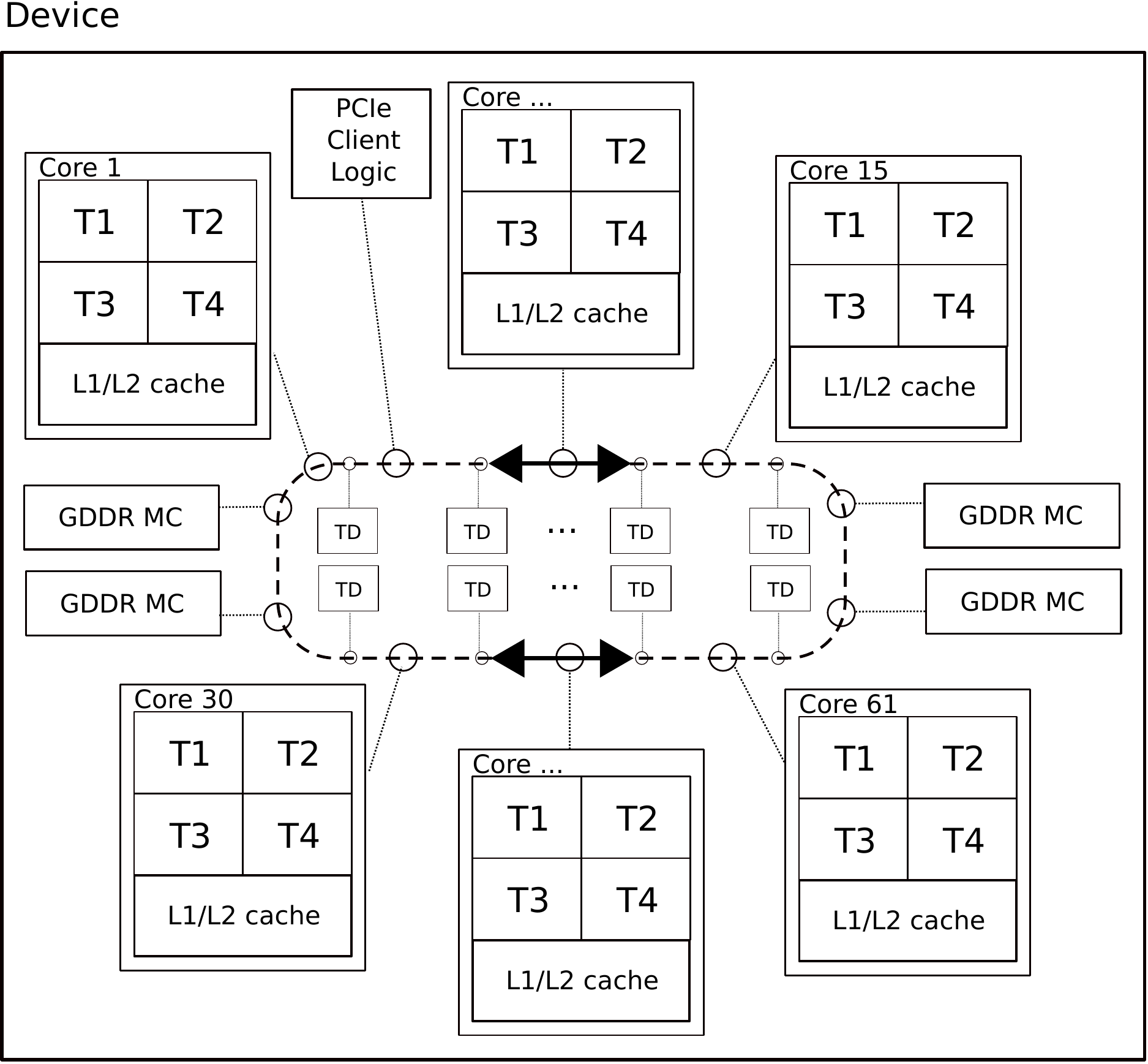}
	\end{center}
	\caption{An overview of our Emil system accelerated with the Intel Xeon Phi.}
	\label{fig:emil-platform}
\end{figure}

Figure \ref{fig:emil-platform} depicts an overview of the Intel Xeon Phi (codenamed Knights Corner) architecture. It is a many-core shared-memory co-processor, which runs a lightweight Linux operating system that offers the possibility to communicate with it over \emph{ssh}. The Xeon Phi offers two programming models:

\begin{enumerate}
	\item \emph{offload} - parts of the applications running on the host are offloaded to the co-processor
	\item \emph{native} - the code is compiled specifically for running natively on the co-processor. The code and all the required libraries should be transferred on the device. In this paper, we focus on the native mode.
\end{enumerate}

The Intel Xeon Phi (type 7120P used in this paper) comprises 61 x86 cores, each core runs at 1.2 GHz base frequency, and up to 1.3GHz on max turbo frequency \cite{chrysos2012intel}. Each core can switch between four hardware threads in a round-robin manner, which amounts to a total of 244 threads per co-processor. Theoretically, the co-processor can deliver up to one teraFLOP/s of double precision performance, or two teraFLOP/s of single precision performance. Each core has its own L1 (32KB) and L2 (512KB) cache. The L2 cache is kept fully coherent by a global distributed tag-directory (TD). The cores are connected through a bidirectional ring bus interconnect, which forms a unified shared L2 cache of 30.5MB. In addition to the cores, there are 16 memory channels that in theory offer a maximum memory bandwidth of 352GB/s.
The GDDR memory controllers provide direct interface to the GDDR5 memory, and the PCIe Client Logic provides direct interface to the PCIe bus. 

Efficient usage of the available vector processing units of the Intel Xeon Phi is essential to fully utilize the performance of the co-processor \cite{TianSPGKMCP13}. Through the 512-bit wide SIMD registers it can perform 16 (16 wide $\times$ 32 bit) single-precision  or 8 (8 wide $\times$ 64 bit) double-precision operations per cycle.

The performance capabilities of the Intel Xeon Phi are discussed and investigated empirically by different researches within several domain applications \cite{dokulil13,lu2013optimizing,DNAxphi,teodoro2013comparative,leung2013investigating,CPE4037}.

\section{Our Parallelization Scheme for Training Convolutional Neural Networks on Intel Xeon Phi}
\label{sec:chaos}

The parallelism can be either divided data-wise, i.e. threads process several inputs concurrently, or model-wise, i.e. several threads share the computational burden of one input. Whether one approach can be advantageous over the other mainly depends on the synchronization overhead of the weight vectors and how well it scales with the number of processing units.

In this section, we first discuss the design aspects of our parallelization scheme for training convolutional neural networks. Thereafter, we discuss the implementation aspects that allow full utilization of the Intel Xeon Phi co-processor.

\subsection{Design Aspects}
\label{sec:chaos-design}

On-line stochastic gradient descent has the advantage of instant updates of weights for each sample. However, the sequential nature of the algorithm yields impediments as the number of multi- and many-core platforms are emerging.
We consider different existing parallelization strategies for stochastic gradient descent: 

\textbf{Strategy A: Hybrid - } uses both data- and model parallelism, such that data parallelism is applied in convolutional layers, and the model parallelism is applied in fully connected layers \cite{krizhevsky2014one}. 

\textbf{Strategy B: Averaged Stochastic Gradient - } divides the input into batches and feeds each batch to a node. This strategy proceeds as follows: (1) Initialize the weights of the learner by randomization; (2) Split the training data into \emph{n} equal chunks and send them to the learners; (3) each learner process the data and calculates the weight gradients for its batch; (4) send the calculated gradients back to the master; (5) the master computes and updates the new weights; and (6) the master sends the new weights to the nodes and a new iteration begins \cite{de2012parallelization}. The convergence speed is slightly worse than for the sequential approach, however the training time is heavily reduced.

\textbf{Strategy C: Delayed Stochastic Gradient - } suggests updating the weight parameters in a round-robin fashion by the workers. One solution is splitting the samples by the number of threads, and let each thread work on its own distinct chunk of samples, only sharing a common weight vector. Threads are only allowed to update the weight vector in a round-robin fashion, and hence each update will be delayed \cite{ZinkevichSL09}.

\textbf{Strategy D: HogWild! - } is a stochastic gradient descent without locks. The approach is applicable for sparse optimization problems (threads/core updates do not conflict much) \cite{recht2011hogwild}.

In this paper, we introduce \textbf{C}ontrolled \textbf{H}ogwild with \textbf{A}rbitrary \textbf{O}rder of \textbf{S}ynchronization (CHAOS), a parallelization scheme that can exploit both thread- and SIMD-level parallelism available on Intel Xeon Phi.  CHAOS is a data-parallel controlled version of HogWild! with delayed updates, which combines parts of strategies A-D. The key aspects of CHAOS are:

\begin{itemize}
	\item \emph{Thread parallelism} - The overview of our parallelization scheme is depicted in Figure \ref{fig:chaos-scheme}. Initially for as many threads as there are available network instances are created, which share weight parameters, whereas to support concurrent processing of images some variables are private to each thread. After the initialization of CNNs and images is done, the process of training starts. The major steps of an epoch include: \emph{Training}, \emph{Validation} and \emph{Testing}. The first step, \emph{Training}, proceeds with each worker picking an image, forward propagates it through the network, calculates the error, and back-propagates the partial derivatives, adjusting the weight parameters. Since each worker picks a new image from the set, other workers do not have to wait for significantly slow workers. After Training, each worker participates in \emph{Validation} and \emph{Testing} evaluating the prediction accuracy of the network by predicting images in the validation and test set accordingly. Adoption of data parallelism was inspired by Krizhevsky \cite{krizhevsky2014one}, promoting data parallelism for convolutional layers as they are computationally intensive.
	
	\item \emph{Controlled HogWild} - during the back-propagation the shared weights are updated after each layer\textquotesingle s computations (a technique inspired by \cite{ZinkevichSL09}), whereas the local weight parameters are updated instantly (a technique inspired by \cite{recht2011hogwild}), which means that the gradients are calculated locally first then shared with other workers. However, the update to the global gradients can be performed at any time, which means that there is no need to wait for other workers to finish their updates. This technique, which we refer to as non-instant updates of weight parameters without significant delay, allows us to avoid unnecessary cache line invalidation and memory writes. 
	
	\item \emph{Arbitrary Order of Synchronization} - There is no need for explicit synchronization, because all workers share weight parameter. However, an implicit synchronization is performed in an arbitrary order because writes are controlled by a first-come-first schedule and reads are performed on demand.
\end{itemize}

\begin{figure}[bt]
	\center
	\includegraphics[width=9cm]{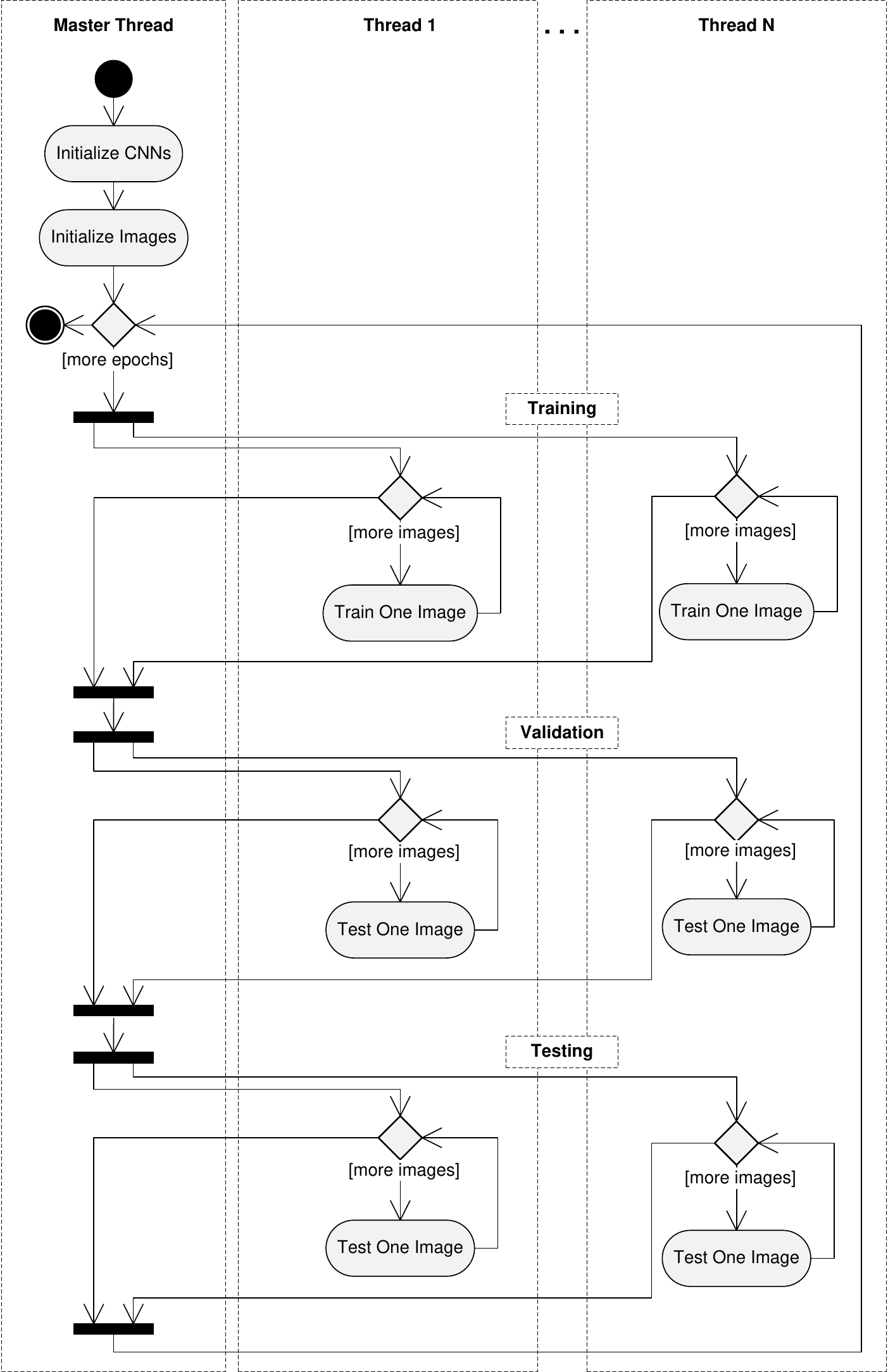}
	\caption{Major activities of \emph{CHAOS} parallelization scheme.}
	\label{fig:chaos-scheme}
	\vspace{-10pt}
\end{figure}

The main goal of CHAOS is to minimize the time spent in the convolutional layers, which can be done through data parallelism, adapting the knowledge presented in strategy A. 
In strategy B, the synchronization is performed because of averaging worker\textquotesingle s gradient calculations. Since work is distributed, computations are performed on stale parameters. The strategy can be applied in distributed and non-distributed settings. The division of work over several distributed workers was adapted in CHAOS.
In strategy C, the updates are postponed using a round-robin-fashion where each thread gets to update when it is its turn. The difference compared to strategy B is that instances train on the same set of weights and no averaging is performed. The advantage is that all instances train on the same weights. The disadvantage of this approach is the delayed updates of the weight parameters as they are performed on stale data. Training on shared weights and delaying the updates are adopted in CHAOS.
Strategy D presents a lock-free approach of updating the weight parameters, updates are performed instantly without any locks. Our updates are not instant, however, after computing the gradients there is nothing prohibiting a worker contributing to the shared weights, the notion of instant inspired CHAOS.

\subsection {Implementation Aspects}
\label{sec:chaos-implementation}

The main goal is to utilize the many cores of the Intel Xeon Phi co-processor efficiently to lower the training time (execution time) of the selected CNN algorithm, at the same time maintaining low deviation in error rates, especially on the test set. Moreover, the quality of the implementation is verified using errors and error rates on the validation and test set.

In the sequential version, only minor modifications of the original version were performed. Mainly, we added a \emph{Reporter} class to serialize execution results. The instrumentation should not add any time penalties in practice. However, if these penalties occur in the sequential version they are likely to imply corresponding penalties in the parallel version, therefore it should not impact the results.

The main goal of the parallel version is to lower the execution time of the sequential implementation and to scale well with the number of processing units on the co-processor. To facilitate this, it is essential to fully consider the characteristics of the underlying hardware. From results derived in the sequential execution we found the hotspots of the application to be predominantly the convolutional layers. The time spent in both forward- and back-propagation is about 94\% of the total time of all layers (up to 99\% for the larger network), which is depicted in the Table \ref{tab:tab_par_fp_conv}.

\begin{table}[t]
	\center
	\caption{Execution times at each layer for the sequential version on the Xeon E5 using the small CNN architecture.
		\label{tab:tab_par_fp_conv}
	}{%
		\begin{tabular}{@{}llll@{}}
			\toprule
			Layer type & Forward propagation & Back-propagation & \% of total \\ 		\midrule
			\textit{Fully connected}  & 40.9 s & 30.9 s  &  1.4 \%  \\  
			\textit{Convolutional}  & 3,241 s & 1,438 s  & 93.7 \%  \\ 
			\textit{Max pooling } & 188.3 s & 8.2 s & 3.9 \%		\\ \bottomrule
		\end{tabular} 
	}
\end{table}

In our proposed strategy, a set of $N$ network instances are created and assigned to $T$ threads. We assume $T == N$, i.e. one thread per network instance. $T$ threads are spawned, each responsible for its own instance. 

The overview of the algorithm is shown in Fig. \ref{fig:chaos-scheme}. In Fig. \ref{fig:train_test_back} the training, testing and back-propagation phase are shown in details. Training (see Fig. \ref{fig:cnn-training}) picks an image, forward propagates it, determines the loss and back-propagates the partial derivatives (deltas) in the network - this process is done simultaneously by all workers, each worker processing one image. Each worker participating in testing (see Fig. \ref{fig:cnn-testing}), picks an image, forward propagates it and then collects errors and error rates. The results are cumulated for all threads. Perhaps the most interesting part is the back-propagation (see Fig. \ref{fig:cnn-backpropagation}). The shared weights are used when propagating the deltas, however, before updating the weight gradients, the pointers are set to the local weights. Thereafter the algorithm proceeds by updating the local weights first. When a worker has contributions to the global weights it can update in a controlled manner, avoiding data races. Updates immediately affect other workers in their training process. Hence the update is delayed slightly, to decrease the invalidation of cache lines, yet almost instant and workers do not have to wait for a longer period before contributing with their knowledge.

\begin{figure}[bt]
	\centering
	\begin{subfigure}[b]{0.22\textwidth}
		\includegraphics[width=\textwidth]{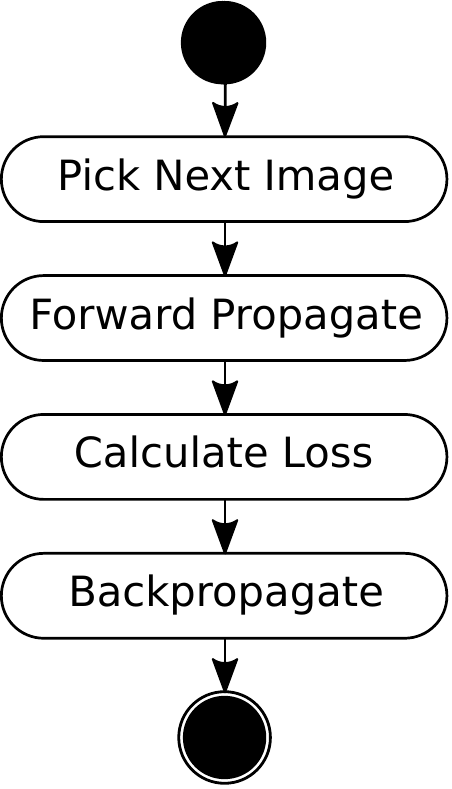}
		\caption{Training}
		\label{fig:cnn-training}
	\end{subfigure}
	~ 
	\begin{subfigure}[b]{0.23\textwidth}
		\includegraphics[width=\textwidth]{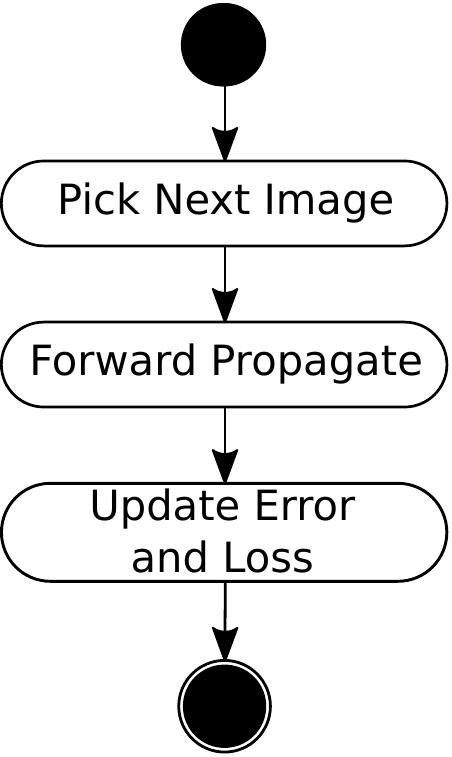}
		\caption{Testing}
		\label{fig:cnn-testing}
	\end{subfigure}
	~ 
	\begin{subfigure}[b]{0.42\textwidth}
		\includegraphics[width=\textwidth]{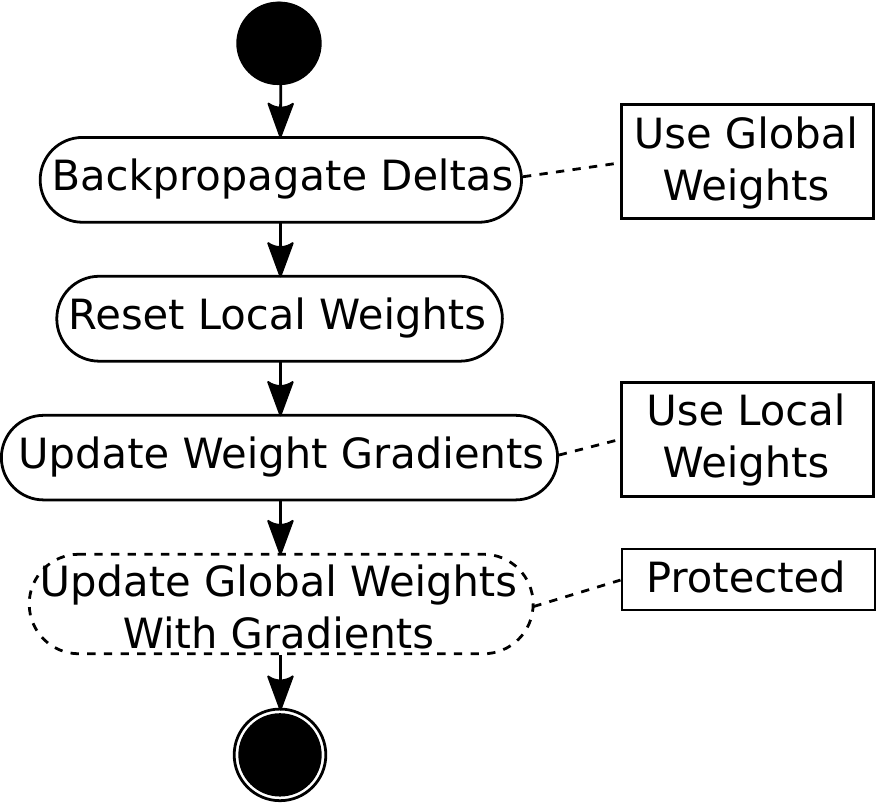}
		\caption{Back-propagation}
		\label{fig:cnn-backpropagation}
	\end{subfigure}
	\caption{The detailed phase of training, testing and back-propagation of one image.}
	\label{fig:train_test_back}
\end{figure}

To see why delays are important, consider the following scenario: If training several network instances concurrently, they share the same weight vectors, other variables are thread private. The major consideration lies in the weight updates. Let $W^j_l$ be the $j$-th weight on the $l$-th layer. In accordance with the current implementation, a weight is updated several times since neurons in a map (on the same layer) share the same weights, and the kernel is shifted over the neurons. Further assume that several threads work on the same weight $W^j_l$ at some point in time. Even if other threads only read the weights, their local data, as saved in the Level 2 cache, will be invalidated and a re-fetch is required to assert their integrity. This happens because cache lines are shared between cores. The approach of slightly delaying the updates and forcing one thread to update in atomicity leads to fewer invalidations. Still a major disadvantages is that the shared weights does not infer any data locality (data cannot retain completely in Level 2 cache for a longer period).

\begin{lstlisting}[frame=single,basicstyle=\scriptsize\ttfamily,breaklines=true,caption={An extract from the vectorization report for the partial derivative updates in the convolutional layer.},label=listing:vectorization]  
remark #15475: --- begin vector loop cost summary ---
remark #15476: scalar loop cost: 30
remark #15477: vector loop cost: 7.500
remark #15478: estimated potential speedup: 3.980
remark #15479: lightweight vector operations: 6
remark #15480: medium-overhead vector operations: 1
remark #15481: heavy-overhead vector operations: 1
remark #15488: --- end vector loop cost summary ---
\end{lstlisting}

To further decrease the time spent in convolutional layers, loops were vectorized to facilitate the vector processing unit of the co-processor. Data was allocated using $\_mm\_malloc()$ with $64$ byte alignment increasing the accuracy of memory requests. The vectorization was achieved by adding $\#pragma~omp~simd$ instructions and explicitly informing the compiler of the memory alignment using $\_\_assume\_aligned()$. Some unnecessary overhead is added through the lack of data alignment of the deltas and weights. 
The computations of partial derivatives and weight gradients in the convolutional layers are performed in a SIMD way, which allows efficient utiliziation of the 512 bit wide vector processing units of the Intel Xeon Phi. An extract from the vectorization report (see Listing \ref{listing:vectorization}), for the updates of partial derivatives in the convolutional layer shows an estimated potential speed up of $3.98\times$ compared to the scalar loop. 

Further algorithmic optimizations were performed. For example: (1) The images are loaded into a pre-allocated memory instead of allocating new memory when requesting an image; (2) Hardware pre-fetching was applied to mitigate the shortcomings of the in-order-execution scheme. Pre-fetching loads data to L2 cache to make it available for future computations; (3) Letting workers pick images instead of assigning images to workers, allow for a smaller overhead at the end of a work-sharing construct; (4) The number of locks are minimized as far as possible; (5) We made most of the variables thread private to achieve data locality.

The training phase was distributed through thread parallelism, dividing the input space over available workers. CHAOS uses the vector processing units to improve performance and tries to retain local variables in local cache as far as possible. The delayed updates decrease the invalidation of cache lines. Since weight parameters are shared among threads, there is a possibility that data can be fetched from another core\textquotesingle s cache instead of main memory, reducing the wait times. Also, the memory was aligned to 64 bytes and unnecessary system calls were removed from the parallel work.

\section{Evaluation}
\label{sec:evaluation}

In this section, we first describe the experimentation environment used for evaluation of our CHAOS parallelization scheme. Thereafter, we describe the development of a performance model for CHAOS. Finally we discuss the obtained results with respect to scalability, speedup, and prediction accuracy. 

\subsection{Experimental Setup}
\label{sec:evaluation-setup}

In this study, OpenMP was selected to facilitate the utilization of thread- and SIMD-parallelism available in the Intel Xeon Phi co-processor. C++ programming language is used for algorithm implementation. The Intel Compiler 15.0.0 was used for native compilation of the application for the co-processor, whereas the $O3$ level was used for optimization.

\emph{System Configuration} - To evaluate our approach we use an Intel Xeon Phi accelerator that comprises 61 cores that run at 1.2 GHz. For evaluation 1, 15, 30, 60, 120, 180, 240, and 244 threads of the co-processor were used. Each thread was responsible for one network instance. For comparison, we use two general purpose CPUs, including the Intel Xeon E5-s695v2 that runs at 2.4 GHz clock frequency, and the Intel Core i5 661 that runs at 3.33GHz clock frequency. 

\emph{Data Set} - To evaluate our approach, the MNIST \cite{lecun2010mnist} dataset of handwritten digits is used. In total the MNIST dataset comprises 70000 images, 60000 of which are used for training/validation and the rest for testing. 

\emph{CNN Architectures} - Three different CNN architectures were used for evaluation, \emph{small, medium} and \emph{large}. The small and medium architecture were trained for 70 epochs, and the large one for 15 epochs, using a starting decay (eta) of 0.001 and factor of 0.9. The small and medium network consist of seven layers in total (one input layer, two convolutional layers, two max-poling layers, one fully connected layer and the output layer). The difference between these two networks is in the number of feature maps per layer and the number of neurons per map. For example, the first convolutional layer of the small network has five feature maps and 3380 neurons, whereas the first convolutional layer of the medium network has 20 feature maps and 13520 neurons. The large network differs from the small and the medium network in the number of layers as well. In total, there are nine layers, one input layer, three convolutional layers, three max-pooling layers, one fully connected layer and the output layer. Detailed information (including the number and the size of feature maps, neurons, the size of the kernels and the weights) about the considered architectures is listed in Table \ref{tab:architectures}.

To address the variability in performance measurements we have repeated the execution of each parallel configuration for three times.

\begin{table}[t]
	\footnotesize
	\center
	\caption{CNN architectures for experimental evaluation of CHAOS.}
	\begin{tabular}{@{}lrrrrrr@{}}
		\toprule 
		& Layer Type & Maps & Map Size & Neurons & Kernel Size & Weights  \\ 
		\midrule 
		\multirow{9}{*}{Large} & Input & - & 29x29 & 841 & - & - \\
		& Convolutional & 20 & 26x26 & 13520 & 4x4 & 340 \\
		& Max-pooling & 20  & 26x26 & 13520 & 1x1 & - \\
		& Convolutional & 60 & 22x22 & 29040  & 5x5 & 30060 \\
		& Max-pooling & 60 & 11x11 & 7260 & 2x2 & -\\
		& Convolutional & 100& 6x6 & 3600 & 6x6 & 216100 \\
		& Max-pooling & 100 & 2x2 & 900 & 3x3 & - \\
		& Fully connected & - & 150  & 150 & - & 135150 \\
		& Output & - & 10 & 10 & - & 1510 \\
		\midrule 
		\multirow{7}{*}{Medium} & Input & - & 29x29 & 841 & - & - \\
		& Convolutional & 20 & 26x26 & 13520 & 4x4 & 340 \\
		& Max-pooling & 20  & 13x13  & 3380 & 2x2 & - \\
		& Convolutional & 40 & 9x9  & 3240 & 5x5 & 20040 \\
		& Max-pooling & 40 & 3x3  & 360 & 3x3 & -\\
		& Fully connected & - & 150 & 150 & - & 54150 \\
		& Output & - & 10 & 10 & - & 1510 \\
		\midrule 
		\multirow{7}{*}{Small} & Input & - & 29x29 & 841 & - & - \\
		& Convolutional & 5 & 26x26 &3380 & 4x4 & 85 \\
		& Max-pooling & 5  & 13x13 & 845 & 2x2 & - \\
		& Convolutional & 10 & 9x9 & 810 & 5x5 & 1260 \\
		& Max-pooling & 10 & 3x3 & 90 & 3x3 & -\\
		& Fully connected & - & 50 & 50 & - & 4550 \\
		& Output & - & 10 & 10 & - & 510 \\
		\bottomrule
	\end{tabular} 
	\label{tab:architectures}
\end{table}

\subsection{Performance Model}
\label{sec:performance-model}

A performance model \cite{perfmod} enables us to reason about the behavior of an implementation in future execution contexts. Our performance model for CHAOS implementation can predict the performance for numbers of threads that go beyond the number of hardware threads supported in the Intel Xeon Phi model that we used for evaluation. Additionally, it can predict the performance of different CNN architectures with various number of images and epochs.

The goal is to construct a parametrized model with the following parameters $ep, i, it$ and $p$, where $ep$ stands for the number of epochs, $i$ indicates the number of images in the training/validation set, $it$ stands for the number of images in the test set, and $p$ is the number of processing units. Table \ref{tab:perf-model-parameters} lists the full set of variables used in our performance model, some of which are hardware dependent and some others are independent of the underlying hardware. Each variable is either measured, calculated, constant, or parameter in the model. Listing \ref{listing:performance-prediction} shows the formula used for our performance prediction model. 

\begin{figure}[bt]
	\captionof{lstlisting}{The formula for our performance prediction model.}
	\begin{mathbox}
		\begin{align}
		T(i,it,ep,p,s) = T_{comp}(i,it,ep,p,s) + T_{mem}(ep,i,p) \notag \\
		= \Bigg(\dfrac{Prep + 4*i +2*it+ 10*ep}{s} \tag{sequential work}\\
		+ \Bigg(\bigg(\Big(\dfrac{FProp+BProp}{s}\Big)*\dfrac{i}{p_i}*ep\bigg) \tag{training}\\
		+ \bigg(\Big(\dfrac{FProp}{s}\Big)*\dfrac{i}{p_i}*ep\bigg) \tag{validation}\\	
		+ \bigg(\Big(\dfrac{FProp}{s}\Big)*\dfrac{it}{p_{it}}*ep\big)\Bigg) \tag{testing}\\
		*CPI\Bigg)*OperationFactor+ T_{mem}(ep,i,p) \notag
		\end{align}
	\end{mathbox}
	\label{listing:performance-prediction}
\end{figure}

The total execution time ($T$) is the sum of computations time ($T_{comp}$) and memory operations ($T_{mem}$). $T$ depends on several factors including: speed, number of processing units, communication costs (such as network latency), and memory contention. The $T_{comp}$ is sum of sequential work, training, validation, and testing. Most interesting is contentions causing wait times, including memory latencies and synchronization overhead. $T_{mem}$ adds memory and synchronization overheads. The contention is measured through an experimental approach by executing a small script on the co-processor for different thread counts, weights and layers. 

\begin{table}[t]
	\center
	\footnotesize
	\caption{Variables used in the performance model.}
	\begin{threeparttable}
		\begin{tabular}{@{}lll@{}}
			\toprule
			Variable & Values & Explanation\\ 
			\midrule
			\multicolumn{3}{c}{Parameters} \\
			\cmidrule(lr){1-3}
			\textit{p} & 1-3,840 & Number of processing units/threads \\ 
			\textit{i} & 60,000&  Number of training/validation images\\ 
			\textit{it} & 10,000 & Number of test images \\ 
			\textit{ep} & 70(small, medium), 15 (large) & Number of epochs \\ 
			\toprule
			\multicolumn{3}{c}{Constants - hardware dependent} \\
			\cmidrule(lr){1-3}
			\textit{CPI} & \begin{tabular}[c]{@{}l@{}}1-2 threads:1\\ 3 threads:1.5\\ 4 threads:2\end{tabular} & Best theoretical CPI/thread \\ 
			\textit{s} & 1.238GHz & Speed of processing unit\\ 
			\textit{OperationFactor} & 15 & Operation factor \\ 
			\toprule
			\multicolumn{3}{c}{Measured - hardware dependent} \\
			\cmidrule(lr){1-3}
			\textit{MemoryContention} & see Table \ref{tab:phi_memory}  & Memory contention \\ 
			\textit{$T_{Fprop}$\tnote{+}} & \begin{tabular}[c]{@{}l@{}}Small: 1.45\\ Medium: 12.55\\ Large: 148.88\end{tabular} & Forward propagation / image (ms)  \\ 
			\textit{$T_{Bprop}$\tnote{+}} & \begin{tabular}[c]{@{}l@{}}Small: 5.3\\ Medium: 69.73\\ Large: 859.19\end{tabular} & Back-propagation / image (ms) \\
			\textit{$T_{Prep}$\tnote{+}} & \begin{tabular}[c]{@{}l@{}}Small: 12.56\\ Medium: 12.7\\ Large: 13.5\end{tabular} & Time for preparations (s) \\  
			\toprule 
			\multicolumn{3}{c}{Calculated - hardware independent} \\
			\cmidrule(lr){1-3}
			\textit{FProp\tnote{*}} & \begin{tabular}[c]{@{}l@{}}Small: 58,000\\ Medium: 559,000\\ Large: 5,349,000\end{tabular} & \# FProp Operations / image \\ 
			\textit{BProp\tnote{*}} & \begin{tabular}[c]{@{}l@{}}Small: 524,000\\ Medium: 6,119,000\\ Large: 73,178,000\end{tabular} & \# BProp Operations / image\\ 
			\textit{Prep\tnote{*}} & \begin{tabular}[c]{@{}l@{}}Small: $10^9$\\ Medium: $10^{10}$\\ Large: $10^{11}$\end{tabular} & \# Operations for preparations \\
			\bottomrule
		\end{tabular}
		\begin{tablenotes}
			{ \footnotesize 
				\item[*] The parameter is only used in prediction a)
				\item[+] The parameter is only used in prediction b)
			}
		\end{tablenotes}
	\end{threeparttable}
	\label{tab:perf-model-parameters}
\end{table}

We define $T_{mem}(ep,i,p) = \frac{MemoryContention * ep *i}{p}$ where $MemoryContention$ is the measured memory contention when $p$ threads are fighting for the I/O weights concurrently. Table \ref{tab:phi_memory} depicts the measured and predicted memory contentions for the Intel Xeon Phi.

Our performance prediction model is not concerned with any practical measurements except for $T_{mem}$. Along with the \textit{CPI} and \textit{OperationFactor} it is possible to derive the number of instructions (theoretically) per cycle that each thread can perform.

\begin{table}[t]
	\center
	\caption{Measured and predicted memory contention for the Intel Xeon Phi.}
	\begin{threeparttable}
		\begin{tabular}{@{}llll@{}}
			\toprule
			\# Threads & Small & Medium & Large\\ 
			\midrule 
			1 & $7.10*10^{-6}$ & $1.56*10^{-4}$ & $8.83*10^{-4}$\\
			15 & $6.40*10^{-4}$ & $2.00*10^{-3}$ & $8.75*10^{-3}$\\ 
			30 & $1.36*10^{-3}$ & $3.97*10^{-3}$ &  $1.67*10^{-2}$ \\ 
			60 & $3.07*10^{-3}$ & $8.03*10^{-3}$ &  $3.22*10^{-2}$ \\ 
			120 & $6.76*10^{-3}$ & $1.65*10^{-2}$ & $6.74*10^{-2}$\\ 
			180 & $9.95*10^{-3}$ & $2.50*10^{-2}$ & $1.00*10^{-1}$\\ 
			240 & $1.40*10^{-2}$ & $3.83*10^{-2}$ &  $1.38*10^{-1}$\\ 
			480\tnote{*} & $2.78*10^{-2}$ & $7.31*10^{-2}$ &	$2.73*10^{-1}$ \\
			960\tnote{*} & $5.60*10^{-2}$ & $1.47*10^{-1}$ & $5.46*10^{-1}$ \\
			1,920\tnote{*} & $1.12*10^{-1}$ & $2.95*10^{-1}$ & $1.09$ \\
			3,840\tnote{*} & $2.25*10^{-1}$ & $5.91*10^{-1}$ &  $2.19$ \\
			\bottomrule
		\end{tabular}
		\begin{tablenotes}
			{ \footnotesize 
				\item[*] Predicted values
			}
		\end{tablenotes}
	\end{threeparttable}
	\label{tab:phi_memory}
\end{table}

We use $Prep$ to be different for each CNN architecture ($10^9, 10^{10}$ and $10^{11}$ for small, medium and large architecture respectively). The $OperationFactor$ is adjusted to closely match the measured value for 15 threads, and mitigate the approximations done for instructions in the first place, at the same time account for vectorization.

When one hardware thread is present per core, one instruction per cycle can be assumed. For 4 threads per core, only 0.5 instructions per cycle can be assumed, which means that each thread gets to execute two instructions every fourth cycle ($CPI$ of 2) and hence we use the $CPI$ factor to control the best theoretical amount of instructions a thread can retire. The speed $s$ is defined in Table \ref{tab:perf-model-parameters}. $FProp$ and $BProp$ are placeholders for the actual number of operations. 

\subsection{Results}
\label{sec:results}

In this section, we analyze the collected data with regards to the execution time and speedup for varying number of threads and CNN architectures. The errors and error rates (incorrect predictions) are used to validate our implementation. Furthermore, we discuss the deviation in number of incorrectly predicted images.

The execution time is the total time the algorithm executes, excluding the time required to initialize the network instances and images (for both the sequential and parallel version). The speed up is measured as the relativeness between two execution times, with the sequential execution times of Intel Xeon E5, Intel Core i5, and Xeon Phi as the base. The error rate is the fraction of images the network was unable to predict and the error the cumulated loss from the loss function.

In the figures and tables in this section, we use the following notations: \emph{Par} refers to the parallel version, \emph{Seq} is the sequential version, and \emph{T} denotes threads, e.g. \emph{Phi Par. 1 T} is the parallel version and one thread on the Xeon Phi.

\textbf{Result 1:} \emph{The CHAOS parallelization scheme scales gracefully to large numbers of threads.}

Figure \ref{fig:result-scalability} depicts the total execution time of the parallel version of the implementation running on the Xeon Phi and the sequential version running on the Xeon E5 CPU. We vary the number of threads on the Xeon Phi between \emph{1, 15, 30, 60, 120, 180, 240,} and \emph{244}, and the CNN architectures between \emph{small, medium} and \emph{large}. We elide the results of \emph{Xeon E5 Seq.} and \emph{Phi Par. 1T} for simplicity and clarity. The large CNN architecture requires 31.1 hours to be completed sequentially on the Xeon E5 CPU, whereas using one thread on the Xeon Phi requires 295.5 hours. By increasing the number of threads to 15, 30, and 60, the execution time decreases to 19.7, 9.9, and 5.0 hours respectively. Using the total number of threads (that is 244) on the Xeon Phi the training may be completed in only 2.9 hours. We may observe a promising scalability while increasing the number of threads. Similar results may be observed for the small and medium architecture.

\begin{figure}[bt]
	\center
	\includegraphics[width=0.8\linewidth]{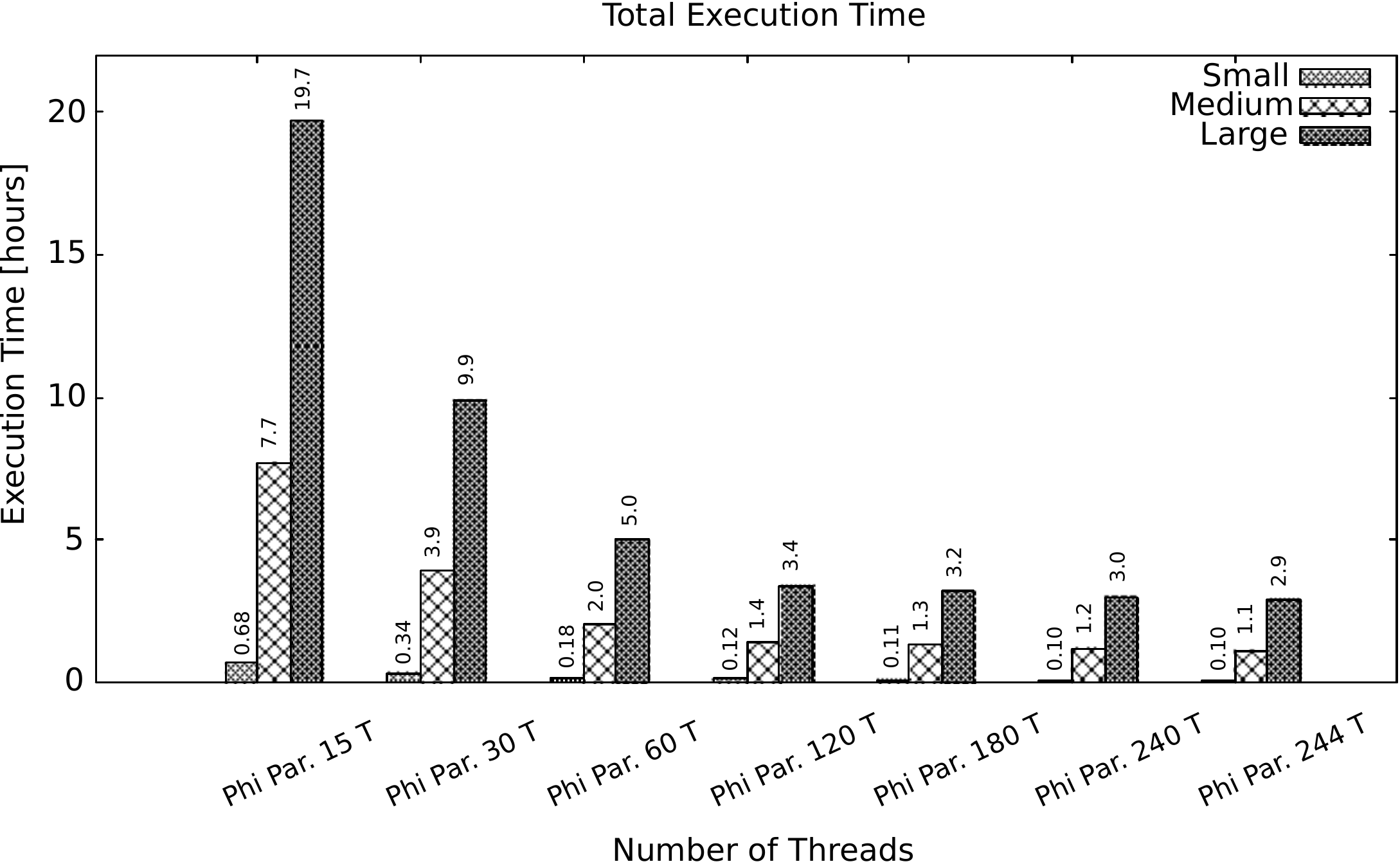}
	\caption{The total execution time for the parallel version executed on the Intel Xeon Phi and the sequential version executed on the Intel Xeon E5.}
	\label{fig:result-scalability}
\end{figure}

It should be considered that the selected CNN architectures were trained for different number of epochs, and that larger networks tend to produce better predictions (lower error rates). A fairer comparison would be to compare the execution times until reaching a specific error rate on the test set. In Fig. \ref{fig:result-scalability-with-stop-criteria} the total execution times for the different CNN architectures and threads on the Xeon Phi is shown. We have set the stop criteria as the \emph{error rate} $\leq 1.54\%$, which is the ending error rate of the test set for the small architecture. The large network executes for a longer period even if it converges in fewer epochs, and that the medium network needs less time to reach an equal (or better) ending error rate than the small and large network. Note that several other factors impact training, including the starting decay, the factor which the decay is decreased, dataset, loss function, preparation of images, initial weight values. Therefore, several combinations of parameters need to be tested before finding a balance. In this study, we focus on the number of epochs as the stop criteria and draw conclusions from this, considering the deviation of the error and error rates.

\begin{figure}[bt]
	\center
	\includegraphics[width=0.8\linewidth]{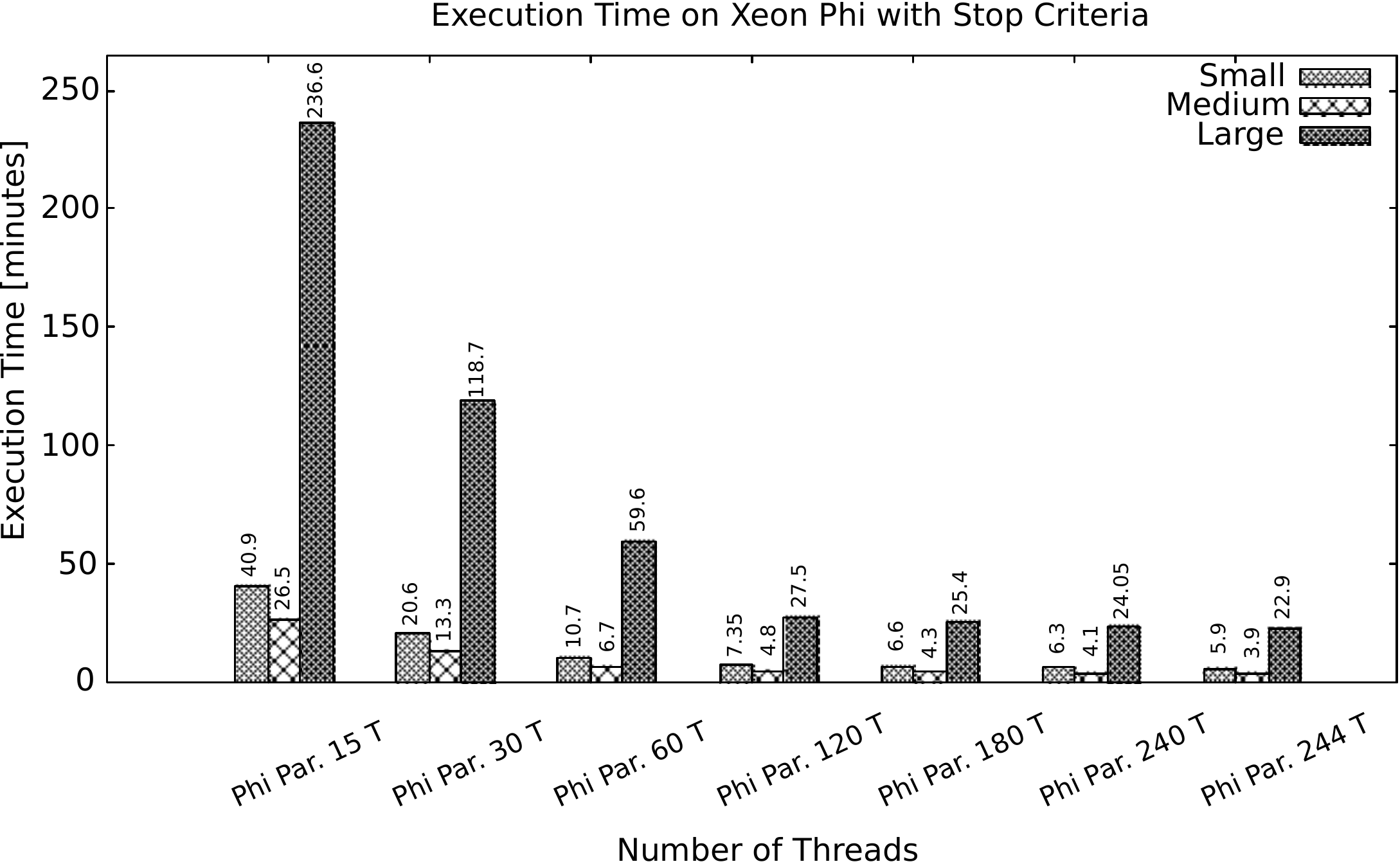}
	\caption{The total execution time for the parallel version executed on the Intel Xeon Phi by setting a stop criteria as the error rate is $\leq 1.54\%$.}
	\label{fig:result-scalability-with-stop-criteria}
\end{figure}

\textbf{Result 2:} \emph{The total execution time is strongly influenced by the forward-propagation and back-propagation in the network. The convolutional layers are the most computationally expensive.}

Table \ref{tab:time-spent-per-layer} depicts the time spent per layer for the large CNN architecture. The results were gathered as the total time spent for all network instances on all layers together. Dividing the total time by the number of network instances and later the number of epochs, yields the number of seconds spent on each layer per network instance and epoch. A lower time spent on each layer per epoch and instance indicates on a speedup.
We may observe that the large architecture spends almost all the time in the convolutional layers and almost no time in the other layers. For \emph{Phi Par. 240 T} about $88\%$ is spent in the back-propagation of convolutional layers and about $10\%$ in forward propagation. We have observed similar results for small and medium CNN architecture, however we elide these results for space. 

We have observed that the more threads involved in training the more percentage of the total time each thread spends in the back-propagation of the convolutional layer, and less time in the others. Overall, the time spent at each layer is decreased per thread when increasing the number of threads. Therefore, there is an interesting relationship between the layer times and the speed up of the algorithm.

\begin{table}[t]
	\centering
	\caption{Average time spent on each layer for the large CNN architecture.}
	\label{tab:time-spent-per-layer}
	\begin{threeparttable}
		\begin{tabular}{@{}lllllllll@{}}
			\toprule
			\multirow{2}{*}{} & \multicolumn{2}{c}{BPF\tnote{1}} & \multicolumn{2}{c}{BPC\tnote{2}} & \multicolumn{2}{c}{FPC\tnote{3}} & \multicolumn{2}{c}{FPF\tnote{4}} \\ 
			& sec          & \%            & sec          & \%            & sec          & \%          	& sec          & \%          \\ \midrule
			\textit{Phi Par. 244 T}    & 7.8          & 1.36\%        & 506.2        & 88.48\%       & 54.7         & 9.56\%        & 0.23         & 0.04\%        \\
			\textit{Phi Par. 240 T}    & 8.1          & 1.34\%        & 532.2        & 88.45\%       & 87.8         & 9.61\%        & 0.24         & 0.04\%        \\
			\textit{Phi Par. 180 T}    & 9.0          & 1.41\%        & 557.9        & 87.78\%       & 64.8         & 10.20\%       & 0.26         & 0.04\%        \\
			\textit{Phi Par. 120 T}    & 11.3         & 1.63\%        & 598.4        & 86.82\%       & 75.4         & 10.94\%       & 0.28         & 0.04\%        \\
			\textit{Phi Par. 60 T}     & 19.5         & 1.91\%        & 877.7        & 86.19\%       & 114.4        & 11.23\%       & 0.47         & 0.05\%        \\
			\textit{Phi Par. 30 T}     & 34.7         & 1.71\%        & 1,749        & 86.36\%       & 228.3        & 11.27\%       & 0.94         & 0.05\%        \\
			\textit{Phi Par. 15 T}     & 60.8         & 1.50\%        & 3,495        & 86.52\%       & 456.9        & 11.31\%       & 1.90         & 0.05\%        \\
			\textit{Phi Par. 1 T}      & 836.7        & 1.38\%        & 52,387       & 86.60\%       & 6,859        & 11.34\%       & 29.75        & 0.05\%        \\ \midrule
			\textit{Xeon E5 Seq.}      & 30.2         & 0.19\%        & 7,097        & 44.51\%       & 8714         & 54.66\%       & 17.04        & 0.11\%        \\ \bottomrule
		\end{tabular}
		\begin{tablenotes}
			{ \footnotesize 
				\item[1] Back-propagation in fully connected layers
				\item[2] Back-propagation of convolutional layers
				\item[3] Forward-propagation of convolutional layers
				\item[4] Forward-propagation in fully connected layers
			}
		\end{tablenotes}
	\end{threeparttable}
\end{table}

Table \ref{tab:speedup-per-layer} presents the speed up relative to the \emph{Phi Par. 1 T} for the different architectures on the convolutional layer. The times are collected by each network instance (through instrumentation of the forward- and back-propagate function) and averaged over the number of network instances and epochs. As can be seen, in almost all cases there is an increase in speed up when increasing the network size, more importantly, the speed up does not decrease. Maybe the most interesting phenomena is that the speed up per layer have an almost direct relationship to the speed up of the algorithm, especially if compared to the back-propagation part. This emphasizes the importance of reducing the time spent in the convolutional layers.

\begin{table}[t]
	\center
	\caption{Averaged layer speed up compared to the \textit{Phi Par. 1 T}.}
	\begin{threeparttable}
		\begin{tabular}{@{}lllllll@{}}
			\toprule
			& BPC-S\tnote{1} & BPC-M\tnote{1} & BPC-L\tnote{1} & FPC-S\tnote{2} & FPC-M\tnote{2}   & FPC-L\tnote{2} \\ \midrule
			\textit{Phi Par. 244 T} & 102.0        & 99.3          & 103.5        & 122.3        & 124.2       & 125.4        \\
			\textit{Phi Par. 240 T} & 96.5         & 94.1          & 98.4         & 114.3        & 117.3       & 118.7        \\
			\textit{Phi Par. 180 T} & 91.8         & 89.5          & 93.9         & 106.3        & 107.0       & 105.8        \\
			\textit{Phi Par. 240 T} & 82.7         & 82.4          & 87.5         & 91.0         & 91.0       & 91.0         \\
			\textit{Phi Par. 60 T}  & 56.9         & 58.9          & 59.7         & 58.6         & 60.1       & 60.0         \\
			\textit{Phi Par. 30 T}  & 29.2         & 29.6          & 29.9         & 29.8         & 30.2        & 30.1         \\
			\textit{Phi Par. 15 T}  & 14.7         & 14.8          & 15.0         & 14.9         & 15.1        & 15.0         \\
			\bottomrule
		\end{tabular}
		\begin{tablenotes}
			{ \footnotesize 
				\item[1]  \textbf{B}ack-\textbf{P}ropagation of \textbf{C}onvolutional layers - \textbf{S}mall, \textbf{M}edium, \textbf{L}arge CNN 
				\item[2] \textbf{F}orward-\textbf{P}ropagation of \textbf{C}onvolutional layers - \textbf{S}mall, \textbf{M}edium, \textbf{L}arge CNN
			}
		\end{tablenotes}
	\end{threeparttable}
	\label{tab:speedup-per-layer}
\end{table}

\textbf{Result 3:} \emph{Using CHAOS parallel implementation for training of CNNs on Intel Xeon Phi we achieved speedups of up to $103\times$, $14\times$, and $58\times$ compared to the single-thread performance on Intel Xeon Phi, Intel Xeon E5 CPU, and Intel Core i5, respectively.}

Figures \ref{fig:result-speedup-vs-E5seq} and \ref{fig:result-speedup-vs-Phi1T} emphasize the facts shown in Fig. \ref{fig:result-scalability} in terms of speedup. Figure \ref{fig:result-speedup-vs-E5seq} depicts the speedup compared to the sequential execution on Xeon E5 (\emph{Xeon E5 Seq.}) for various number of threads and CNN architectures. As can be seen, adding more threads results in speedup increase in all cases. Using 240 threads on the Xeon Phi infer a $13.26\times$ speedup for the small CNN architecture. Utilizing the last core of the Xeon Phi, which is used by the OS, shows even higher speedup ($14.07\times$). We may observe that doubling the number of threads from 15, to 30, and from 30 to 60 almost doubles the speedup (2.03, 4.03, and 7.78). Increasing the number of threads further results with significant speedup, but the double speedup trend breaks.

\begin{figure}[bt]
	\center
	\includegraphics[width=0.8\linewidth]{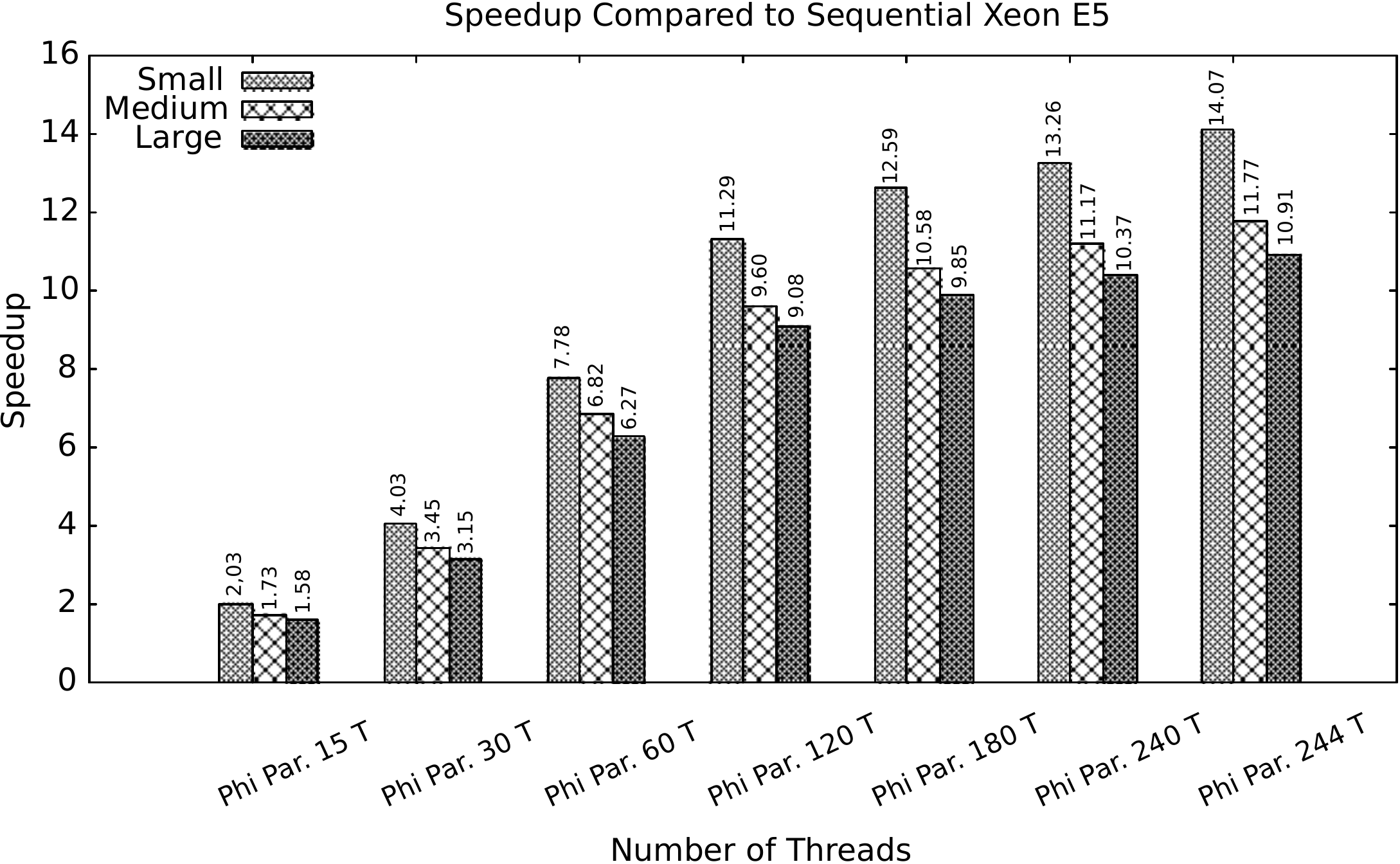}
	\caption{Speedup of the three CNN architectures by varying the number of threads compared to the sequential execution on Intel Xeon E5.}
	\label{fig:result-speedup-vs-E5seq}
\end{figure}

Figure \ref{fig:result-speedup-vs-Phi1T} shows the speedup compared to the execution running in one thread of the Xeon Phi (\emph{Phi Par. 1 T}) while varying the number of threads and the CNN architectures. We may observe that the speedup is close to linear for up to 60 threads for all CNN architectures. Increasing the number of threads further results with significant speedup. Moreover it can be seen that when keeping the number of threads fixed and increasing the architecture size, the speed up increases with a small factor as well, except for 244 threads. It seems like larger architectures are beneficial. However, it could also be the case that \emph{Phi Par. 1 T} executes relatively slower than \emph{Xeon E5 Seq.} for larger architectures than for smaller ones.

\begin{figure}[bt]
	\center
	\includegraphics[width=0.8\linewidth]{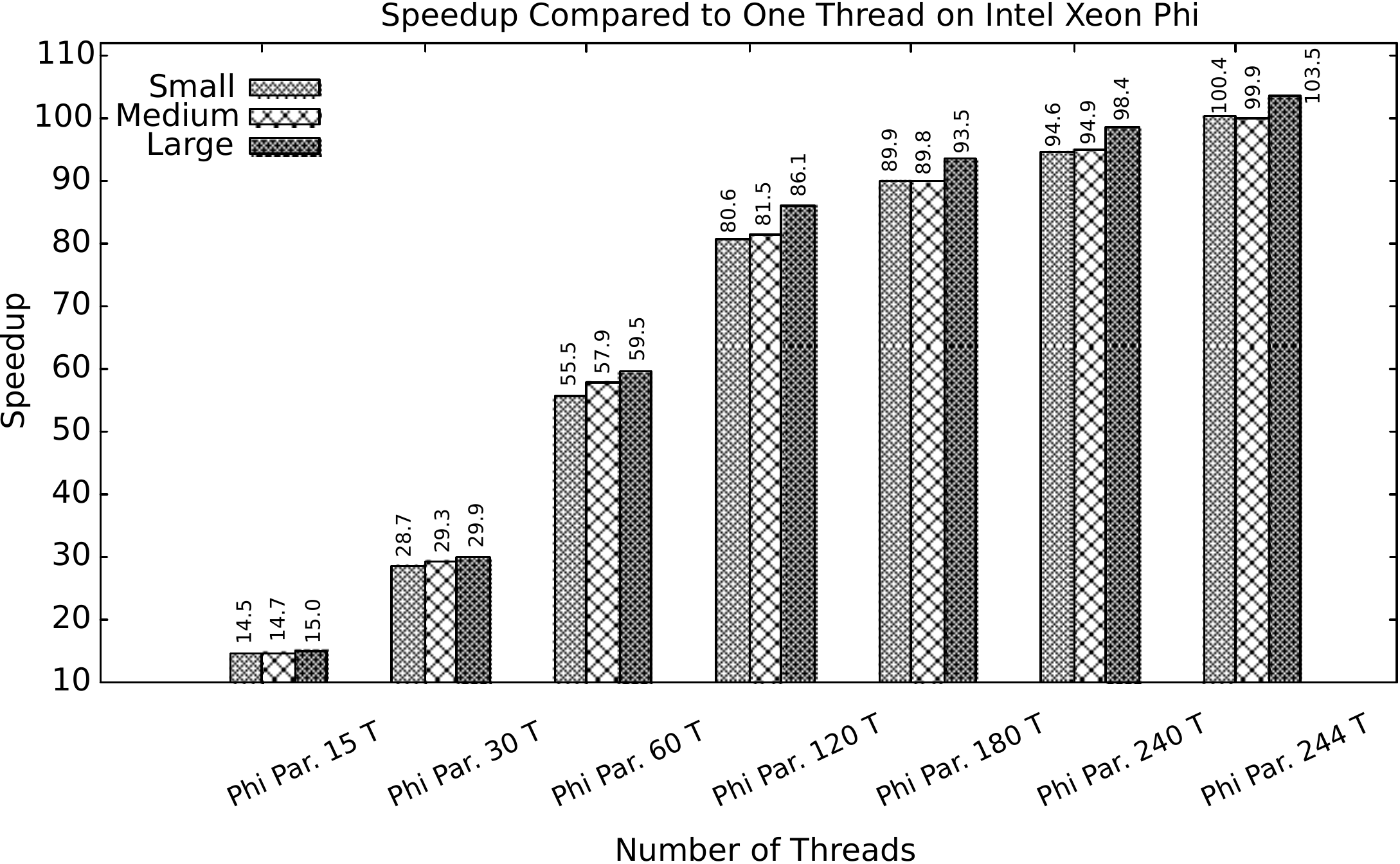}
	\caption{Speedup of the three CNN architectures by varying the number of threads compared to one thread on Intel Xeon Phi.}
	\label{fig:result-speedup-vs-Phi1T}
\end{figure}

Figure \ref{fig:result-speedup-vs-i5Seq} shows the speedup compared to the sequential version executed in Intel Core i5 (\emph{Core i5 Seq.}) while varying the number of threads and the CNN architectures. We may observe that using 15 threads we gain $10\times$ speedup. Doubling the number of threads to 30, and then to 60 results with close to double speedup increase (19.8 and 38.3). By using 120 threads (that is two threads per core) the trend of double speedup increase breaks ($55.6\times$). Increasing the number of threads per core to three and four results with modest speedup increase ($62\times$ and $65.3\times$). 

\begin{figure}[bt]
	\center
	\includegraphics[width=0.8\linewidth]{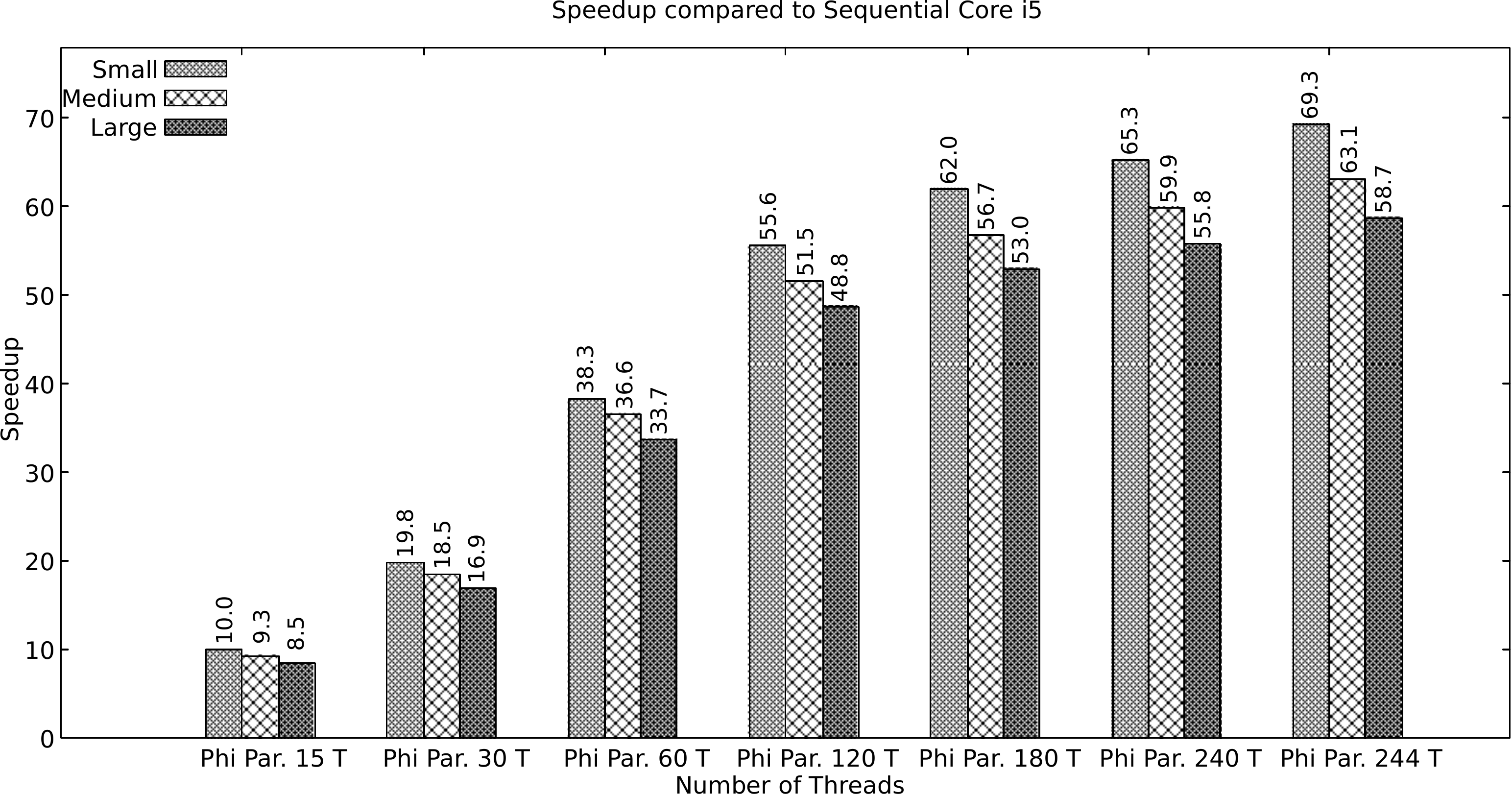}
	\caption{Speedup of the three CNN architectures by varying the number of threads compared to one thread on Intel Core i5.}
	\label{fig:result-speedup-vs-i5Seq}
\end{figure}

\textbf{Result 4:} \emph{The image classification accuracy of parallel implementation using CHAOS is comparable to the one running sequentially. The deviation error and the number of incorrectly predicted images is not abundant.}

We validate the implementation by comparing the error and error rates for each epoch and configuration. Figure \ref{fig:prediction-accuracy} depicts the ending errors for the three considered CNN architectures for both validation and test set. The black dashed line delineates the base line (that is a ratio of 1). Values below the line are considered better, whereas those above the line are worse than for \emph{Xeon E5 Seq}. As a base line, we use the Xeon E5, however identical results are derived executing the sequential version on any platform. As can be seen in Fig. \ref{fig:prediction-accuracy}, the largest difference is encountered by \emph{Phi Par. 244 T}, about 22 units (0.05\%) worse than the base line. On the contrary, \emph{Phi Par. 15 T} has 9 units\textquotesingle lower error compared to the base line for the large test set. The validation sets are rather stable whereas the test set fluctuates more heavily. Although one should consider the deviation in error respectfully, they are not abundant in this case. Please note that the diagram has a high zoom factor, hence the differences are magnified.

\begin{figure}[bt]
	\center
	\includegraphics[width=0.8\linewidth]{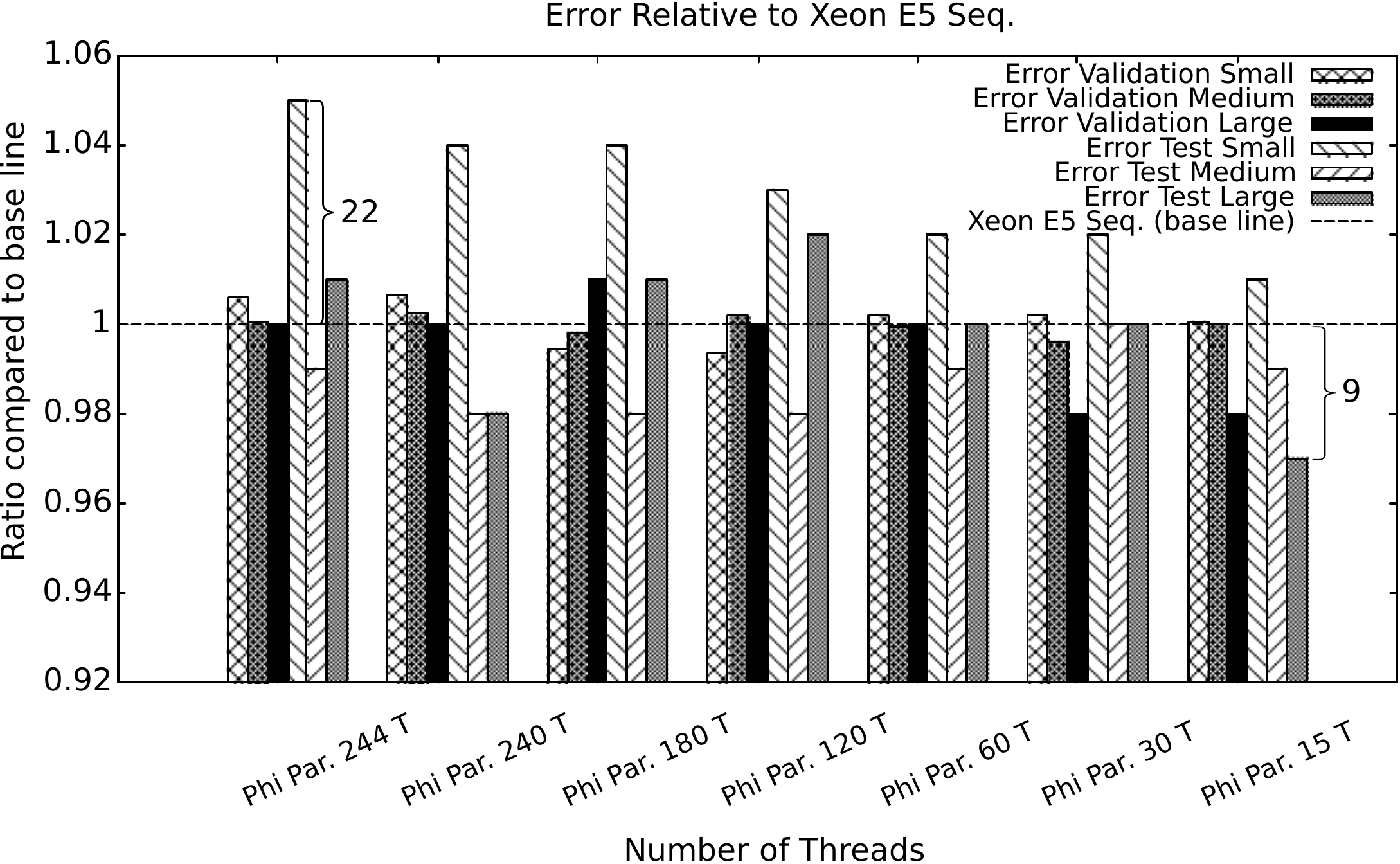}
	\caption{The relative cumulative error (loss) for the three considered CNN architectures (small, medium, and large) for both validation and test set.}
	\label{fig:prediction-accuracy}
\end{figure}

Table \ref{tab:err_images} lists the number of incorrectly classified images for each CNN architecture. For each architecture, the total (\emph{Tot}) number of images and the difference (\emph{Diff}) compared to the optimal numbers of \emph{Xeon E5 Seq.} are shown. Negative values indicate that the ending error rate was better than optimal (less images were incorrectly predicted), whereas positive values indicate that more images than Xeon E5 Seq. were incorrectly predicted. For each column in the table, \emph{best} and \emph{worst} values are annotated with \emph{underline} and \emph{bold} fonts, respectively. No obvious pattern can be found, however, increasing the number of threads does not lead to worse prediction in general. \emph{Phi Par. 180 T} stands out as it was 17 images better than \emph{Xeon E5 Seq.} for small architecture on validation set. \emph{Phi Par. 15 T} also performs worst on the small architecture on the validation set. The overall worst performance is achieved by \emph{Phi par. 120 T} on the test set for small CNN architecture. Please note that the total number of images in the validation set is $60,000$ and $10,000$ for the test set. Overall, the number of incorrectly predicted images and the deviation from the base line is not abundant.

\begin{table}[t]
	\small
	\center
	\caption{The number of incorrectly classified images for different CNN architectures.}
	\begin{tabular}{@{}c|llllll|llllll@{}}
		\toprule
		\# & \multicolumn{6}{c}{Validation} &  \multicolumn{6}{c}{Test} \\
		of Phi & \multicolumn{2}{c}{Small} & \multicolumn{2}{c}{Medium} & \multicolumn{2}{c}{Large} & \multicolumn{2}{c}{Small} & \multicolumn{2}{c}{Medium} & \multicolumn{2}{c}{Large}  \\
		threads & Tot & Diff & Tot & Diff & Tot & Diff & Tot & Diff & Tot & Diff & Tot & Diff \\ \midrule
		\textit{244} & 616                        & 4             & 85                          & 1             & 12                         & 2             & 155                        & 2              & 98                          & 3              & \textbf{95}                & \textbf{1}     \\
		\textit{240} & 610                        & -2            & 86                          & 2             & 11                         & 1             & 154                        & 1              & \underline{95}                 & \underline{0}     & 91                         & -3             \\
		\textit{180} & \underline{595}               & \underline{-17}  & \textbf{87}                 & \textbf{3}    & \textbf{12}                & \textbf{2}    & 158                        & 5              & 98                          & 3              & 95                         & 1              \\
		\textit{120} & 607                        & -5            & 83                          & -1            & 11                         & 1             & \textbf{159}               & \textbf{6}     & 95                          & 0              & 94                         & 0              \\
		\textit{60}  & 615                        & 3             & \underline{81}                 & \underline{-3}   & 11                         & 1             & 156                        & 3              & 98                          & 3              & 91                         & -3             \\
		\textit{30}  & 612                        & 0             & 83                          & -1            & \underline{10}                & \underline{0}    & 156                        & 3              & 98                          & 3              & 90                         & -5             \\
		\textit{15}  & \textbf{617}               & \textbf{5}    & 84                          & 0             & \underline{10}                & \underline{0}    & \underline{153}               & \underline{0}     & \textbf{100}                & \textbf{5}     & \underline{84}                & \underline{-10}    \\ \bottomrule
	\end{tabular}
	\label{tab:err_images}
\end{table}

\textbf{Result 5:} \emph{The predicted execution times obtained from the performance model match well the measured execution times.}

Figures \ref{fig:pred-model-small}, \ref{fig:pred-model-medium}, and \ref{fig:pred-model-large} depict the predicted and measured execution times for small, medium and large CNN architecture. For the small network (see Fig. \ref{fig:pred-model-small}), the predictions are close to the measured values with a slight deviation at the end. The prediction model seems to over-estimate the execution time with a small factor. 

\begin{figure}[bt]
	\center
	\includegraphics[width=0.8\linewidth]{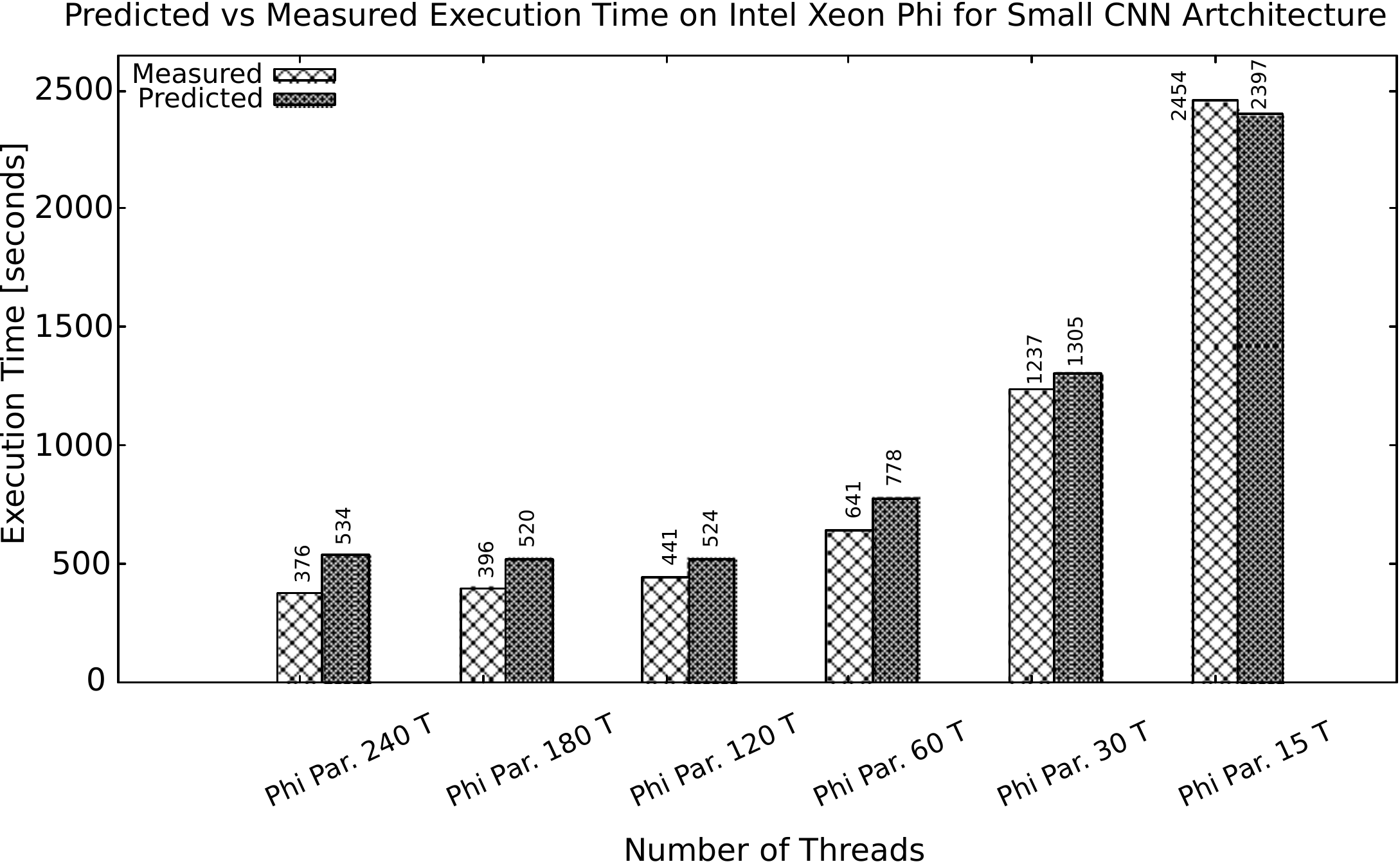}
	\caption{The comparison between the predicted execution time and the measured execution time on Intel Xeon Phi for the small CNN architecture.}
	\label{fig:pred-model-small}
\end{figure}

For the medium architecture (see Fig. \ref{fig:pred-model-medium}) the prediction follow the measured values closely, although it underestimates the execution time slightly. At 120 threads, the measured and predicted values starts to deviate, which are recovered at 240 threads.

\begin{figure}[bt]
	\center
	\includegraphics[width=0.8\linewidth]{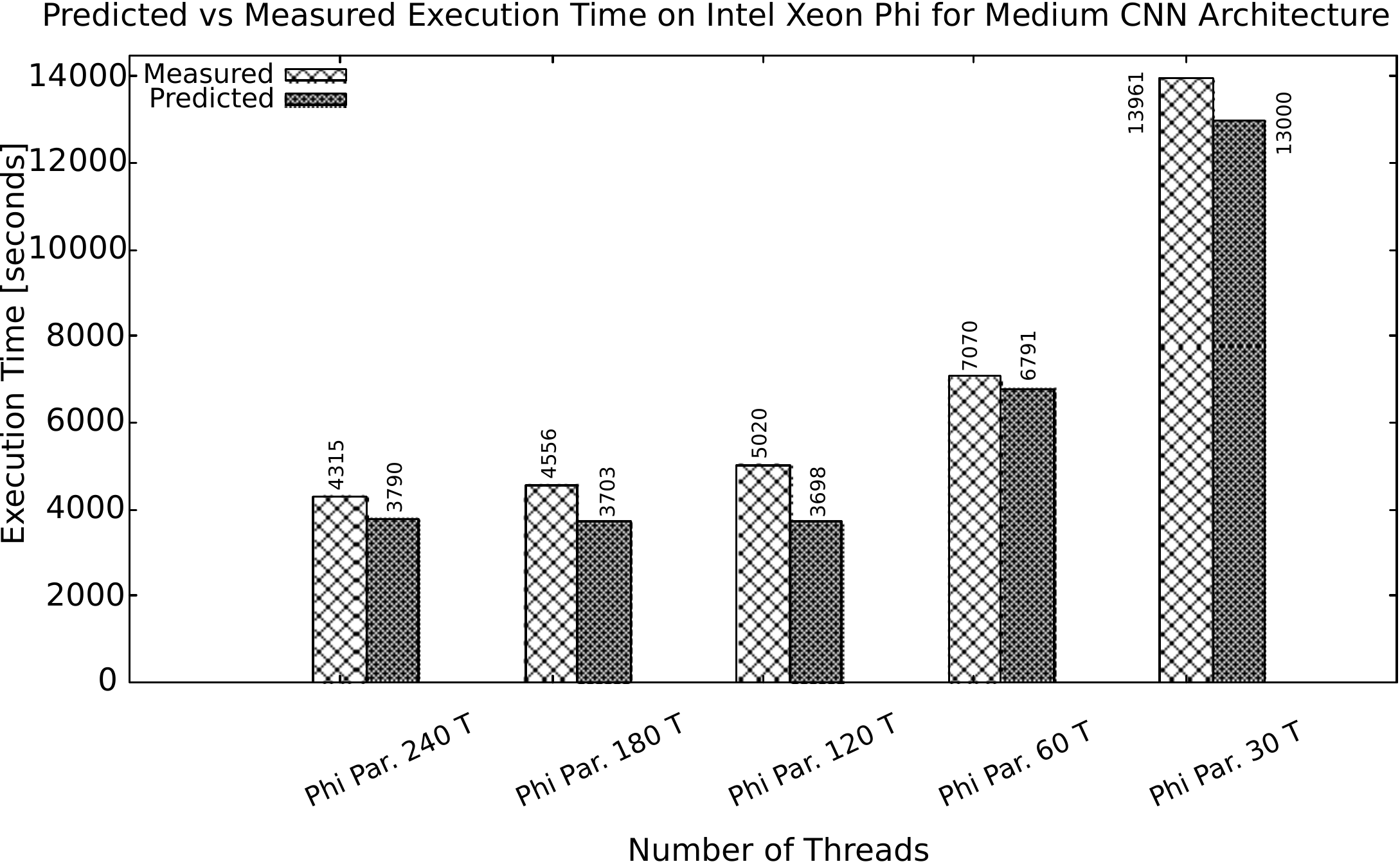}
	\caption{The comparison between the predicted execution time and the measured execution time on Intel Xeon Phi for the medium CNN architecture.}
	\label{fig:pred-model-medium}
\end{figure}

The large architecture yields similar results as the medium. As can be seen, the measured values are slightly higher than the predictions, however, the predictions follow the measured values. As can be seen for 120 threads there is a deviation which is recovered for 240 threads. Also, the predictions increase between 120 and 180, and 180 and 240 threads for both predictions, whereas the actual execution time is lowered. This is most probably due to the CPI factor that is added when 3 or more threads are present on the same core. We use the expression $x = \dfrac{\abs{m-p}}{p}$ to calculate the deviation in predictions for our prediction model and all considered architectures, where $m$ is the measured and $p$ is the predicted value. The average deviations over all measured thread counts are as follows: $14.57\%$ for the small CNN, $14.76\%$ for medium, and $15.36\%$ for large CNN.

\begin{figure}[bt]
	\center
	\includegraphics[width=0.8\linewidth]{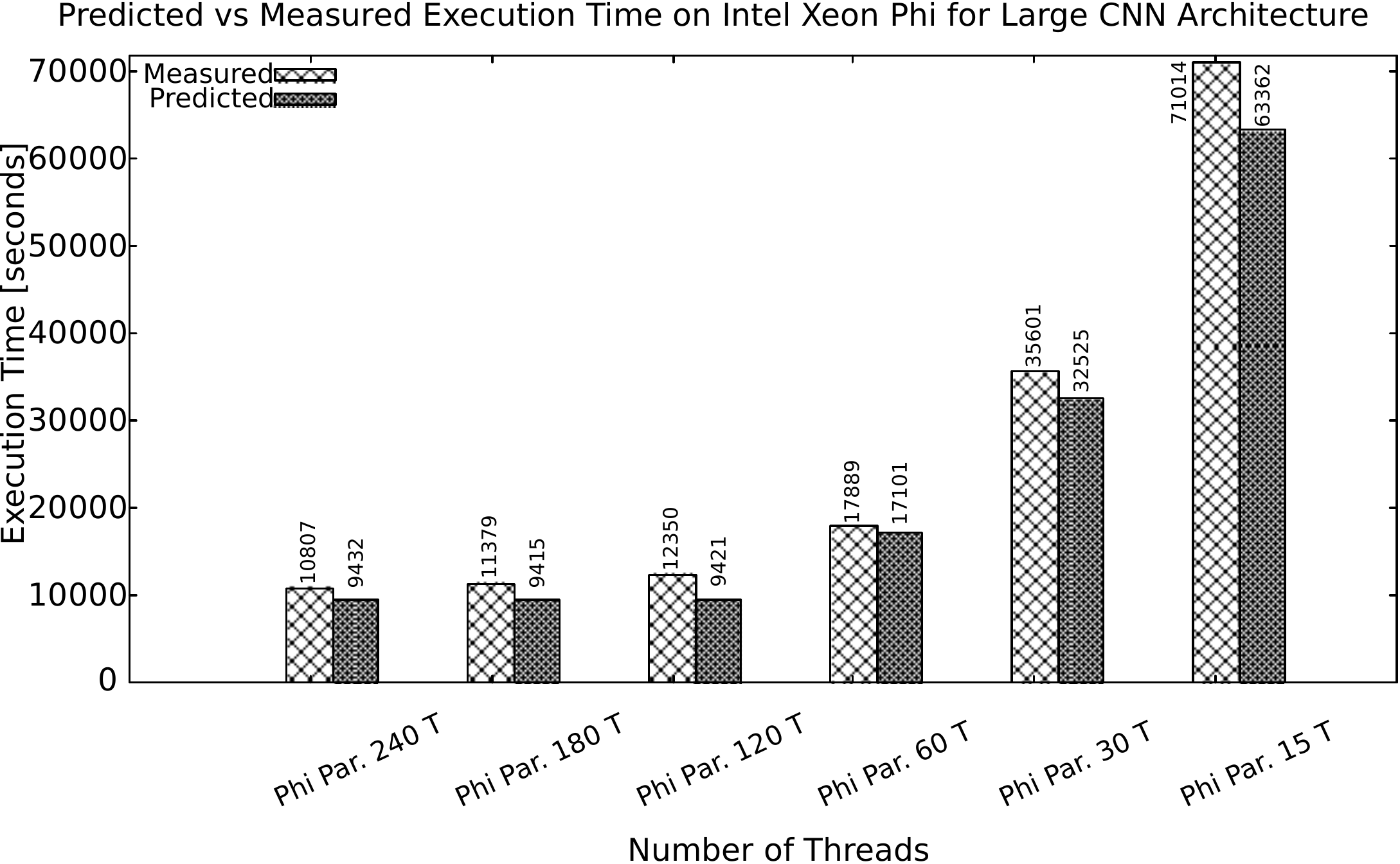}
	\caption{The comparison between the predicted execution time and the measured execution time on Intel Xeon Phi for the large CNN architecture.}
	\label{fig:pred-model-large}
\end{figure}

\textbf{Result 6:} \textit{Prediction of execution time for number of threads that go beyond the 240 hardware threads of the model of Intel Xeon Phi used in this paper show that CHAOS scales well up to several thousands of threads.}

We used the prediction model to predict the execution times for 480, 960, 1920, and 3840 threads for different CNN architectures, using the same parameters. The results in Table \ref{tab:pred-model-results} show that if \textit{3,840} threads were available, the small network should take about \textit{4.6} minutes to train, the medium \textit{14.5} minutes and the large \textit{36.8} minutes. The predictions for the large CNN architecture are not as well aligned when increasing to larger thread counts as for small and medium. 

\begin{table}[t]
	\center
	\caption{Predicted execution times (min) for 480, 960, 1,920 and 3,840 threads using the performance models.}
	\begin{tabular}{@{}lllll@{}}
		\toprule
		\# Threads & \multicolumn{1}{c}{480} & \multicolumn{1}{c}{960} & \multicolumn{1}{c}{1,920} & \multicolumn{1}{c}{3,840} \\ 
		\midrule
		\textit{Small CNN}  & 6.6  & 5.4  & 4.9  & 4.6 \\ 
		\textit{Medium CNN} & 36.8 & 23.9 & 17.4 & 14.2 \\
		\textit{Large CNN} 	& 92.9 & 60.8 & 44.8 & 36.8 \\ \bottomrule
	\end{tabular}
	\label{tab:pred-model-results}
\end{table}

Additionally, we evaluated the execution time for varying image counts, and epochs, for 240 and 480 threads for the small CNN architecture. As can be seen in Table \ref{tab:futurepredictions} doubling the number of images or epochs, approximately doubles the execution time. However, doubling the number of threads does not reduce the execution time in half.

\begin{table}[t]
	\centering
	\caption{The execution times in minutes when scaling epochs and images for 240 and 480 threads using the performance model on the small CNN architecture. }
	\begin{threeparttable}
		\begin{tabular}{@{}llllllllll@{}}
			\toprule
			&                        & \multicolumn{4}{c}{240 Threads} & \multicolumn{4}{c}{480 Threads} \\ \midrule
			\multicolumn{2}{c}{Images}              & \multicolumn{4}{c}{Epochs}      & \multicolumn{4}{c}{Epochs}      \\ 
			\multicolumn{1}{c}{$i$\tnote{1}} & \multicolumn{1}{c}{$it$\tnote{2}} & 70    & 140   & 280    & 560    & 70    & 140   & 280    & 560    \\ \midrule
			\multicolumn{1}{c}{60k}                    & 10k                     & 8.9   & 17.6  & 35.0   & 69.7   & 6.6   & 12.9  & 25.6   & 51.1   \\
			\multicolumn{1}{c}{120k}                   & 20k                     & 17.6  & 35.0  & 69.7   & 139.3  & 12.9  & 25.6  & 51.1   & 101.9  \\
			\multicolumn{1}{c}{240k}                   & 40k                     & 35.0  & 69.7  & 139.3  & 278.3  & 25.6  & 51.1  & 101.9  & 203.6  \\ \bottomrule
		\end{tabular}
		\begin{tablenotes}
			{ \footnotesize 
				\item[1] Number of images in the training/validation set
				\item[2] Number of images in the test set
			}
		\end{tablenotes}
	\end{threeparttable}
	\label{tab:futurepredictions}
\end{table}

\section{Summary and Future Work}
\label{sec:conclusion}
	
Deep learning is important for many modern applications, such as, voice recognition, face recognition, autonomous cars, precision medicine, or computer vision. We have presented CHAOS that is a parallelization scheme to speed up the training process of Convolutional Neural Networks. CHAOS can exploit both thread- and SIMD-parallelism of Intel Xeon Phi co-processor. Moreover, we have described our performance prediction model, which we use to evaluate our parallelization solution and infer the performance on future architectures of the Intel Xeon Phi. Major observations include,

\begin{itemize}
	\item CHAOS parallel implementation scales well with the increase of the number of threads;
	\item convolutional layers are the most computationally expensive part of the CNN training effort; for instance, for 240 threads, 88\% of the time is spent on the back-propagation of convolutional layers;
	\item using CHAOS for training CNNs on Intel Xeon Phi we achieved up to $103\times$, $14\times$, and $58\times$ speedup compared to the single-thread performance on Intel Xeon Phi, Intel Xeon E5 CPU, and Intel Core i5, respectively;	
 	\item image classification accuracy of CHAOS parallel implementation is comparable to the one running sequentially;
	\item predicted execution times values obtained from our performance model match well the measured execution times;
	\item results of the performance model indicate that CHAOS scales well beyond the 240 hardware threads of the Intel Xeon Phi that is used in this paper for experimentation.
\end{itemize}

Future work will extend CHAOS to enable the use of all cores of host CPUs and the co-processor(s).


\end{document}